\documentclass[aps,pre,floats,floatfix,showpacs,superscriptaddress,nofootinbib,twocolumn]{revtex4}
\usepackage{amsmath,amssymb,multirow,amsfonts}
\usepackage[pdftex]{graphicx}
\usepackage[tight]{subfigure}
\usepackage{color}

\def\ie{\rm{i.e.\ }}
\def\eg{\rm{e.g.\ }}
\def\P{\mathbb{P}}

\begin{document}

\title{Hamiltonian Dynamics of Preferential Attachment}

\author{Konstantin Zuev}
\affiliation{Department of Physics, Northeastern University, Boston, MA 02115, USA}
\author{Fragkiskos Papadopoulos}
\affiliation{Department of Electrical Engineering, Computer Engineering and Informatics, Cyprus University of Technology, 33 Saripolou Street, 3036 Limassol, Cyprus}
\author{Dmitri Krioukov}
\affiliation{Department of Physics, Department of Mathematics, Department of Electrical\&Computer Engineering, Northeastern University, Boston, MA 02115, USA}

\begin{abstract}

Prediction and control of network dynamics are grand-challenge problems in network science. The lack of understanding of fundamental laws driving the dynamics of networks is among the reasons why many practical problems of great significance remain unsolved for decades. Here we study the dynamics of networks evolving according to preferential attachment, known to approximate well the large-scale growth dynamics of a variety of real networks. We show that this dynamics is Hamiltonian, thus casting the study of complex networks dynamics to the powerful canonical formalism, in which the time evolution of a dynamical system is described by Hamilton's equations. We derive the explicit form of the Hamiltonian that governs network growth in preferential attachment. This Hamiltonian turns out to be nearly identical to graph energy in the configuration model, which shows that the ensemble of random graphs generated by preferential attachment is nearly identical to the ensemble of random graphs with scale-free degree distributions. In other words, preferential attachment generates nothing but random graphs with power-law degree distribution. The extension of the developed canonical formalism for network analysis to richer geometric network models with non-degenerate groups of symmetries may eventually lead to a system of equations describing network dynamics at small scales.

\end{abstract}


\maketitle

\section{Introduction}

Large real networks---social, biological, or technological---are complex dynamical systems~\cite{Newman10-book,Kleinberg10-book,Dorogovtsev10-book}. Understanding the dynamics of these systems is a key to better prediction and control of their behavior, and the behavior of the processes running on them, such as epidemic spreading~\cite{MeAr09,PSV2001} and cascading failure propagation~\cite{LiBaBuStHa12,LiYiKaSh14}. Network dynamics can be roughly split into two categories: large-scale and small-scale. Large-scale dynamics usually means network growth, for example the growth of the Internet over years. Small-scale dynamics refers to the dynamics of links in a given network at small time scales, for example real-time interactions among mobile phone users or genes in a cell. It is quite unlikely that the small-scale dynamics of different networks can be in any way similar, so it seems that natural options to study and predict this dynamics can only be purely phenomenological, including data mining, model building, and parameter fitting~\cite{Ko09,JaCl14}, with all their caveats~\cite{owhadi2015,BuAn02}.
However if two dynamical systems behave differently, it does not mean that the laws that govern their dynamics are different---the simplest example would be the quite different dynamics of two and three gravitating bodies of similar masses in empty space. And indeed if considering network dynamics we move from the small to large scale, we observe that preferential attachment~\cite{BarAlb99,KraReLe00,DoMeSa00} accurately describes the growth of many very different real networks~\cite{Ne01,BaJeNe02,VaPaVe02a,JeNeBa03}. This observation raises two questions: (1)~can preferential attachment be formulated within the canonical approach in physics, and if so, then (2)~does the same approach apply to network dynamics at small scales?


Here we answer positively the first question by showing that preferential attachment can be fully described within the canonical formalism. That is, we show that the growth dynamics of networks evolving according to preferential attachment is Hamiltonian, and derive the explicit form of the Hamiltonian in the corresponding Hamilton's equations. This Hamiltonian turns out to be nearly identical to the Hamiltonian in the soft configuration model.

In the canonical formalism the dynamics of a system with canonical coordinates $(q,p)$ and Hamiltonian $\mathcal{H}(q,p,t)$, which is usually the total energy of the system, is described by Hamilton's equations
\begin{equation}
\label{eq:HamEqintro}
\dot{q}=\frac{\partial\mathcal{H}}{\partial p},\hspace{10mm} \dot{p}=-\frac{\partial\mathcal{H}}{\partial q}.
\end{equation}
The canonical approach has a long history of success in physics. All the fundamental interactions in nature are described by Euler--Lagrange or Hamilton's equations with different symmetry groups~\cite{RyderQFT1996}. The Einstein field equations in general relativity are Euler--Lagrange equations for the gravitational Einstein--Hilbert action, while the ADM formalism is the corresponding Hamiltonian formulation~\cite{Wald2010}. Here we extend the canonical formalism to complex networks, and find the Hamiltonian describing the dynamics of growing networks in preferential attachment.

Most network models can be classified as either equilibrium models, which are the ensembles of graphs of fixed size, \eg the Erd\H{o}s--R\'{e}nyi random graphs~\cite{SolRap,ErRe59,ErRe60}, or non-equilibrium models, in which networks grow with time, \eg the preferential attachment model~\cite{BarAlb99,KraReLe00,DoMeSa00}. Equilibrium models are usually more amendable for analytical treatment, while non-equilibrium models better mimic the growth dynamics of real networks. In the special case of uncorrelated random graphs, there exist growing network models that produce equilibrium ensembles of graphs with an arbitrary degree distribution~\cite{DoMeSa03b}. In general however, the two types of models, growing and equilibrium, are very different, and so are the ensembles of random graphs that they define. A great number of works have studied equilibrium ensembles. A statistical theory of equilibrium correlated random graphs was developed in~\cite{BL02}. Review article~\cite{Fetal04} surveys important advances in the field of equilibrium network models. In particular, it discusses the structural properties of the graphs and topological phase transitions in equilibrium graph ensembles. An interesting ``symbiotic'' model, an equilibrium network model with fixed number of nodes and links which evolves using a local rewiring move, was studied in~\cite{BM03}. It was shown that if the graph Hamiltonian is chosen appropriately, then the networks generated by the model are scale-free. In all the prior works however, the graph Hamiltonians appear only in the equilibrium sense, \ie as the graph energy proportional to the logarithm of the graph probability in the equilibrium ensemble. To the best of our knowledge no prior work has studied Hamiltonian dynamics of growing networks, where the graph Hamiltonian is the graph energy which defines the dynamics of growing graphs via Hamilton's equations~(\ref{eq:HamEqintro}).

Our starting point relies on recent results~\cite{KrOs13} establishing the special conditions under which there exists strong equivalence or duality between equilibrium and non-equilibrium (growing) graph ensembles $(\mathcal{G}_N,\mathbb{P})$, where $\mathcal{G}_N$ is a set of graphs of size~$N$, and $\mathbb{P}$ is a probability distribution on $\mathcal{G}_N$. Two network models or graph ensembles $(\mathcal{G}_N,\mathbb{P}_1)$ and $(\mathcal{G}_N,\mathbb{P}_2)$ are equivalent if they generate any graph $G\in\mathcal{G}_N$ with the same probability, \ie $\mathbb{P}_1(G)=\mathbb{P}_2(G)$ for any~$G$. We say that two ensembles are strongly equivalent, if they are equivalent for any~$N$. Using these special conditions, we obtain an intermediate result, which is important in its own right. It gives an equilibrium formulation of preferential attachment. This formulation is useful because it allows, for the first time to the best of our knowledge, to explicitly calculate for any given graph $G$, e.g., a given real network, the probability $\mathbb{P}(G)\propto\exp[-H(G)]$ with which preferential attachment generates this graph, where $H(G)$ is the graph Hamiltonian (graph energy). This Hamiltonian turns out to be very similar to the Hamiltonian in the soft configuration model~\cite{PaNe04,GaLo08,AnBi09}. Based on this intermediate result, we then derive the dynamic Hamiltonian $\mathcal{H}$ that governs the network evolution in preferential attachment.

Remarkably, the static Hamiltonian $H$ and its dynamic counterpart $\mathcal{H}$ turn out to be nearly identical.
The only difference between the two is that exact node degrees in $H$ are replaced by their expected values in $\mathcal{H}$. We thus prove that preferential attachment and configuration model are in fact the same ensemble, or in other words, that preferential attachment generates nothing but random graphs with a given power-law degree distribution. One could in principle expect that to be true in view of several equilibrium(-like) approaches to preferential attachment~\cite{BoPa03,BiBu05,Berger2014PA}.
In~\cite{BoPa03} it is shown how the hidden variable formalism introduced for equilibrium graph ensembles can be applied to the preferential attachment networks to derive their degree distribution and the correlation structure. In a similar spirit, \cite{BiBu05} applies methods of statistical mechanics to compare equilibrium and non-equilibrium graph ensembles and, in particular, demonstrates that the degree distribution in preferential attachment is identical to that in the equilibrium ensemble of random trees. Recent work~\cite{Berger2014PA}, among other results, proves that a minor modification of the preferential attachment model, called sequential preferential attachment, is identical to the equilibrium graph ensemble constructed using several P\'{o}lya urn processes. In this work we prove that the expectation that preferential attachment and configuration model are nearly equivalent is indeed correct.

The flow of logic in the paper, and a more detailed summary of the results are as follows. We begin with a recollection of basic facts concerning exponential random graph models~(ERGMs)~\cite{HoLe81,FS86,PaNe04,Ko09} (Section~\ref{sec:ERGs}) and models of random graphs with hidden variables~\cite{BoPa03,CaCaDeMu02} (Section~\ref{sec:equivalence}) that we will need in subsequent sections. In particular, in latter models, an equilibrium random graph ensemble is fully defined by a distribution~$\rho(r)$ of hidden variables~$r$ attached to nodes, and connection probability~$p(r,r')$ between nodes, and two such ensembles are equivalent ($\P_1(G)=\P_2(G)$), as soon their hidden variable distributions and connection probabilities are the same, $\rho_1(r)=\rho_2(r)$ and $p_1(r,r')=p_2(r,r')$. Random graphs with hidden variables are not ERGs {\it per se}, but they are collections of ERGMs with fixed values of hidden variables playing the role of Lagrange multipliers.

We then briefly discuss soft preferential attachment (SPA, Section~\ref{sec:SPA}). SPA is different from standard
preferential attachment (PA) in only that in PA new links attach to
existing node $i$ with probability proportional to its degree $k_i$,
while in SPA new links attach to existing node $i$ with probability
proportional to its \emph{expected} degree $\kappa_i$. The idea behind the SPA definition is to assign to each new node $j$
a hidden variable $r_j \sim \log j$, and then connect $j$ to existing node
$i$ with certain probability that depends only on the current values
of $j$'s and $i$'s hidden variables $r_j$ and $r_i$. The words \emph{current
values} appear here because the values of these hidden variables~$r$
need to be updated as the network grows, and the combination of
this update rule and connection probability are such that new nodes
do indeed connect to existing nodes with probability proportional
to their current expected degrees, so that we do have SPA. The
key point behind SPA is that it has a coupled dynamics of hidden
variables $r_i$ and expected degrees $\kappa_i$, both growing functions
of the network size $N=1,2,\ldots$ or ``cosmological time'' $t \sim \log N$.

Next, in Section~\ref{sec:PAasERG}, we show that SPA is an ERGM, which is asymptotically ($N\gg1$) identical to the soft configuration model (SCM) for sparse graphs with average degree $\bar{k}\ll N$.
We do this in steps. We first recall, in Section~\ref{sec:SCM}, that the SCM is an ERGM
with Hamiltonian $H(G) = \sum_i k_i r_i + C$, where $k_i$ is the
degree of node $i$ in graph $G$, $r_i$ is the Lagrange multiplier
fixing the expected values of $k_i$ in this ERG ensemble to some
value $\langle k_i \rangle$, and $C$ are some additional terms. This ensemble is an
equilibrium ensemble of random graphs with a given sequence of
expected degrees $\langle k_i \rangle$. If this sequence is power-law-distributed,
then the sequence of $r_i$s is exponentially distributed.
If this $r$-sequence is not fixed but sampled, for each graph, from
a fixed exponential distribution~$\rho(r)$, then the resulting SCM, is an ERGM with hidden variables~$r$ of random graphs
with a given expected power-law degree distribution.
The next two most technical sections deal with certain cosmetic
adjustments to SCM (SCM$^+$, Section~\ref{sec:SCM+}) and SPA ($\widetilde{\mathrm{SPA}}$, Section~\ref{sec:tildeSPA}),
and in Section~\ref{sec:tSPA=SCM+} we show that after these adjustments, the two models (SCM$^+$ and $\widetilde{\mathrm{SPA}}$) are asymptotically equivalent,
and derive their equilibrium ERG Hamiltonians~$H$ in Section~\ref{sec:all-erg-hamiltonians}.

Finally, in Section~\ref{sec:HamDynSPA}, we turn back to the dynamics of SPA,
and pose the question:
is there a dynamic Hamiltonian $\mathcal{H}(\kappa_i,r_i,t)$ such that the
dynamics of expected degrees $\kappa_i(t)$ and hidden variables $r_i(t)$ in SPA is the solution of
Hamilton's equation with this Hamiltonian~$\mathcal{H}$? We answer this
question positively by first deriving the exact dynamics of $\kappa_i$s in SPA and $\widetilde{\mathrm{SPA}}$ (Sections~\ref{sec: kappa SPA} and~\ref{sec: kappa tSPA}), and then finding a whole family of
Hamiltonians that provide a solution to the question above (Section~\ref{sec:dynamic-hamiltonians}).
This family is parameterized by arbitrary functions $\xi_i(t)$, and we show in the same section
that $\xi_i(t)$ can be selected such that the resulting dynamic
Hamiltonian $\mathcal{H}$ in $\widetilde{\mathrm{SPA}}$, evaluated on the solution of Hamilton's equation, is equal,
for any value of graph size $N$, to the equilibrium SCM$^+$ Hamiltonian~$H$ in Section~\ref{sec:all-erg-hamiltonians},
upon substitution $\kappa_i(N) = k_i$.

\section{Exponential Random Graphs}
\label{sec:ERGs}

The exponential random graph model (ERGM) \cite{HoLe81,FS86,PaNe04,Ko09} is one of the most popular and well-studied equilibrium network models, also known as the $p^*$ model in the social network research community \cite{WaPa96,AnWaCr99,RoPaKaLu07}. ERGM is a graph ensemble $(\mathcal{G}_N,\mathbb{P})$, where $\mathcal{G}_N$ is the set of all simple graphs (\ie undirected graphs without self-loops or multi-edges) on $N$ nodes, and $\mathbb{P}$ is the probability distribution on $\mathcal{G}_N$ that maximizes the Gibbs entropy
\begin{equation}\label{eq:entropy}
S(\mathbb{P})=-\sum_{G\in\mathcal{G}_N}\mathbb{P}(G)\ln\mathbb{P}(G)\rightarrow\max,
\end{equation}
subject to the constraints
\begin{equation}\label{eq:constrants}
\langle x_i\rangle =\bar{x}_i, \hspace{3mm} i=1,\ldots,r.
\end{equation}
The $x_i$ in the above relation are certain graph properties (\eg number of edges or number of triangles) often referred to as the graph ``observables'',  $\bar{x}_i$ are the prescribed expected values of these observables in the model, and $\langle\cdot\rangle$ denotes the expectation with respect to $\mathbb{P}$. Intuitively, an ERGM  is a ``maximally random'' ensemble of graphs with fixed values $\bar{x}_i$ for certain ensemble averages $\langle x_i\rangle$. Mathematically, the maximization of randomness corresponds to the maximization of the entropy (\ref{eq:entropy}).   Constraining the expected rather than exact values of graph observables relaxes the topological conditions on the network and makes the model amenable to analytical treatment.

The constrained optimization problem (\ref{eq:entropy}) and (\ref{eq:constrants}) can be solved by the standard method of Lagrange multipliers, and it has the following explicit solution \cite{PaNe04}
\begin{equation}\label{eq: P=exp(H)/Z}
\mathbb{P}(G)=\frac{e^{-H(G)}}{Z},
\end{equation}
where
\begin{equation}
Z=\sum_{G\in\mathcal{G}_N} e^{-H(G)}
\end{equation}
is the partition function, \ie the normalizing constant for distribution (\ref{eq: P=exp(H)/Z}), and
\begin{equation}
H(G)=\sum_{i=1}^r{\theta_ix_i(G)}
\end{equation}
is the graph Hamiltonian, \ie the energy of microstate $G$ in the equilibrium Boltzmann distribution  (\ref{eq: P=exp(H)/Z}). The parameters $\theta_i$ are the Lagrange multipliers (``auxiliary fields'') coupled to observables $x_i$. They are determined by the following system of $r$ equations
\begin{equation}\label{eq:df/dtheta=xbar}
\frac{\partial F}{\partial\theta_i}=\bar{x}_i, \hspace{3mm} i=1,\ldots,r,
\end{equation}
where $F=-\ln Z$ is the free energy. The ERGM distribution (\ref{eq: P=exp(H)/Z}) is thus fully determined by the observables $x_i$ and their expected values $\bar{x}_i$. In~\cite{ZuEiKr15}, the ERGM is extended to exponential random simplicial complexes.

As an example of an ERGM, which  we will refer to in Section~\ref{sec:SCM}, consider the edge-independent random graph (EIRG) model. In this case, the graph observables are the graph edges: $x_{ij}=a_{ij}$, where $a=(a_{ij})$ is the adjacency matrix of $G\in\mathcal{G}_N$. The constrains (\ref{eq:constrants}) are then
\begin{equation}
\langle a_{ij}\rangle=p_{ij}, \hspace{3mm} i<j, \hspace{2mm} i,j=1,\ldots,N,
\end{equation}
where $0\leq p_{ij}\leq1$,
and the Hamiltonian is
\begin{equation}
H(G)=\sum_{i<j}\theta_{ij}a_{ij}.
\end{equation}
Unlike many other examples, the partition function $Z$ can be calculated exactly for the EIRG model~\cite{PaNe04}:
\begin{equation}
Z=\prod_{i<j}\left(1+e^{-\theta_{ij}}\right).
\end{equation}
The relationship between the Lagrange multipliers $\theta_{ij}$ and the model parameters $p_{ij}$ follows from (\ref{eq:df/dtheta=xbar})
\begin{equation}\label{eq:fermi-dirac}
p_{ij}=\frac{1}{1+e^{\theta_{ij}}}.
\end{equation}
Knowing the partition function allows to find the corresponding ERGM distribution (\ref{eq: P=exp(H)/Z})
\begin{equation}\label{eq:P_EIRG}
\mathbb{P}(G)=\prod_{i<j}p_{ij}^{a_{ij}}(1-p_{ij})^{1-a_{ij}}.
\end{equation}
This expression immediately suggests how to generate graphs from the maximum-entropy ensemble ($\mathcal{G}_N,\mathbb{P}$): connect every pair $(i,j)$ of distinct nodes $i\neq j$, $i,j=1,\ldots,N$, independently at random with probability $p_{ij}$. We remark that (\ref{eq:fermi-dirac}) is nothing but the Fermi-Dirac distribution, where the Lagrange multiplier $\theta_{ij}$ is interpreted as the energy of the ``single-particle'' state $(i,j)$. Throughout the paper, we will often use the so-called classical limit for the Fermi-Dirac distribution, \ie if the energy  $\theta_{ij}$ is large, then $p_{ij}\approx e^{-\theta_{ij}}$.

	\section{Random Graphs with Hidden Variables}\label{sec:equivalence}

	Random graphs with hidden variables~\cite{BoPa03,CaCaDeMu02} are ensembles of random graphs in which graphs are generated (or sampled) as follows.
Each node $i=1,\ldots,N$ is first assigned a hidden variable $r_i$, sampled from the probability distribution $\rho(r)$, and then each pair of nodes $(i,j)$ is connected with probability $p_{ij}=p(r_i,r_j)$. Since the hidden random variables are independent, the probability $\mathbb{P}(G)$ of graph $G$ in the ensemble $(\mathcal{G}_N,\mathbb{P})$ is
	\begin{equation}\label{eq:P(G)eqiens}
	\begin{split}
	\mathbb{P}(G)&=\int\mathbb{P}(G|\mathbf{r})\rho(\mathbf{r})d\mathbf{r}\\
	&=\int\prod_{i>j}p_{ij}^{a_{ij}}(1-p_{ij})^{1-a_{ij}}\prod_{i=1}^N\rho(r_i)dr_i,
	\end{split}
	\end{equation}
	where $(a_{ij})$ is $G$'s adjacency matrix and $\mathbf{r}=(r_1,\ldots,r_N)$. This equilibrium graph ensemble is thus fully defined by two functions: the hidden variable PDF $\rho(r)$ and the connection probability function $p(r,r')$.
	
	In equilibrium statistical mechanics, two ensembles are equivalent if their state occupation probabilities are the same for any state. Similarly, two graph ensembles $(\mathcal{G}_N,\mathbb{P}_1)$ and $(\mathcal{G}_N,\mathbb{P}_2)$ are equivalent if they generate any graph $G\in\mathcal{G}_N$ with the same probability:
\begin{equation}
\mathbb{P}_1(G)=\mathbb{P}_2(G).
\end{equation}
It follows from Eq.~(\ref{eq:P(G)eqiens}) that two ensembles of random graphs with hidden variables are equivalent if their hidden variable distributions and connection probabilities are the same:
\begin{align}
\rho_1(r) &= \rho_2(r),\label{eq:same-rho}\\
p_1(r,r') &= p_2(r,r').\label{eq:same-p}
\end{align}

Random graphs with hidden variables are closely related to exponential random graphs. If we sample all hidden variables $r_i$ from $\rho(r)$ just once, and then fix them, then the probability of graph $G$ in the ensemble with these fixed $r_i$s is given by (\ref{eq:P_EIRG}), with $p_{ij}=p(r_i,r_j)$. Rewriting these $p_{ij}$s as $1/\left(1+e^{\theta_{ij}}\right)$ makes the ensemble manifestly identical to the EIRG ensemble with Lagrange multipliers $\theta_{ij}$. Therefore one can think of graphs with unfixed (sampled) hidden variables as a collection EIRGs with ``randomized'' Lagrange multipliers sampled from a fixed distribution.

\section{Soft Preferential Attachment}\label{sec:SPA}

Our first goal is to represent the preferential attachment (PA) model as an ERGM. The original formulation of PA \cite{BarAlb99}, where a new node connects to an existing node with probability proportional to its degree, is very intuitive, but not convenient for this  purpose. Instead, we will use a hidden variable formulation of PA.
It was shown in \cite{PKSBK12} that PA can be formulated as a hidden variable model, which generates growing networks up to some size $N$ with average degree $\bar{k}$ and power-law exponent $\gamma \geq 2$, as follows. For each new node $i=1,\ldots,N$:
\begin{enumerate}
	\item Assign to node $i$ hidden variable
    \begin{equation}
    r_i=\ln i.
    \end{equation}
	\item Update the values of hidden variables of all existing nodes $j<i$ by setting
    \begin{eqnarray}
    r_j(i)&=&\beta r_j+(1-\beta)r_i\text{, where}\\
    \beta&=&\frac{1}{\gamma-1}.
    \end{eqnarray}
	\item Connect node $i$ to each existing node $j<i$ with probability
	\begin{equation}
	p_{ij}=\frac{1}{1+e^{r_j(i)+r_i-R_i}},
	\end{equation}
	where
	\begin{equation}
	R_i=r_i-\ln\frac{1-e^{-(1-\beta)r_i}}{m(1-\beta)} \hspace{2mm} \mbox{and} \hspace{2mm} m=\frac{\bar{k}}{2}.
	\end{equation}
\end{enumerate}

In large sparse networks, where $r_i$ is large and $\beta < 1$, 
the linking probability
\begin{equation}
p_{ij}\approx m(1-\beta)e^{-r_j(i)} \propto e^{-r_j(i)}=i^{-1}\left(\frac{i}{j}\right)^\beta.
\end{equation}
In Section~\ref{sec: kappa SPA}, we show that the expected degree of node $j$ at time $i$ is
\begin{equation}
\label{eq:bar_k_1}
\kappa_j(i)=\frac{m(1-\beta)}{\beta}\left(\frac{i}{j}\right)^\beta+\frac{m(2\beta-1)}{\beta},
\end{equation}
and therefore, the probability that $i$ connects to $j$ is approximately a liner function of $j$'s \textit{expected} degree $\kappa_j(i)$. We note that Eq.~(\ref{eq:bar_k_1}) holds for $\beta < 1$, i.e., $\gamma > 2$. The corresponding relation for the limit $\beta \to 1$ ($\gamma \to 2$) is derived in Appendix~\ref{A1}.  We refer to the hidden variable formulation of PA as the \textit{soft} preferential attachment (SPA) model. Figure~\ref{Fig1}(a) shows a doubly logarithmic plot of the empirical degree distribution in a network generated by SPA along with the fitted power-law distribution.

A conceptually similar formulation of preferential attachment as an equilibrium network model with hidden variables was first introduced in~\cite{BoPa03}, where the hidden variable $r_i$ of node $i$ is simply its injection time, $r_i=i$. For technical reasons that will become apparent in the next section, here we define $r_i=\ln i$, so that $r_i$ can be identified with the cosmological time of birth of node~$i$~\cite{KrPaKi10,PKSBK12}.

\section{SPA as an ERGM}\label{sec:PAasERG}

Let $(\mathcal{G}_N,\mathbb{P}_{\mathrm{SPA}})$ be the ensemble of graphs induced by SPA, where $\mathbb{P}_{\mathrm{SPA}}(G)$ is the probability that SPA generates $G\in\mathcal{G}_N$. In this section we show that
\begin{equation}\label{eq:P_spa propto}
\mathbb{P}_{\mathrm{SPA}}(G)\propto e^{-H_{\mathrm{SPA}}(G)},
\end{equation}
where the SPA Hamiltonian $H_{\mathrm{SPA}}$ is intimately related to the Hamiltonian in the soft configuration model.

\subsection{Soft Configuration Model} \label{sec:SCM}

The soft configuration model (SCM)~\cite{PaNe04,GaLo08,AnBi09} is an ERGM, where graph observables are node degrees $k_i$, $i=1,\ldots,N$. The model has various equivalent formulations, and, in particular, SCM appears as a special degenerate case of the equilibrium hyperbolic model \cite{KrPaKi10} in a certain limiting parameter regime. This formulation also belongs to the wide class of network models with hidden variables. Specifically, the SCM formulation in \cite{KrPaKi10} generates equilibrium networks of size $N$ with average degree $\bar{k}$ and power-law exponent $\gamma > 2$, as follows:
\begin{enumerate}
	\item For $i=1,\ldots,N$, assign to node $i$ hidden variable
	\begin{equation}
	r_i\sim\rho_{\mathrm{SCM}}(r)\approx \alpha e^{\alpha(r-R_{\mathrm{SCM}})}, \hspace{3mm} 0\leq r\leq R_{\mathrm{SCM}},
	\end{equation}
	where $\alpha=\beta^{-1}=\gamma-1$  and
	\begin{equation}
	R_{\mathrm{SCM}}=\ln\frac{N}{\bar{k}(1-\beta)^2}.
	\end{equation}
	\item Connect nodes $i$ and $j$, $j\neq i$, with probability
	\begin{equation}\label{eq:scm p_ij}
	p_{ij}=\frac{1}{1+e^{r_i+r_j-R_{\mathrm{SCM}}}}.
	\end{equation}
\end{enumerate}

A more familiar (but less convenient for our purposes) formulation of the SCM is obtained by the change of hidden variables $h=h_0e^{R_{\mathrm{SCM}}-r}$, $h_0=\bar{k}(1-\beta)$. The hidden variable $h$ has then the power-law distribution $h\sim\rho(h)\propto h^{-\gamma}$ and the connection probability is $p_{ij}=1/(1+N\bar{k}/h_ih_j)$~\cite{SqGa11,CoBo12,KrOs13}.
Comparing the connection probabilities in EIRG (\ref{eq:fermi-dirac}) and SCM (\ref{eq:scm p_ij}), we readily obtain that the Lagrange multiplier $\theta_{ij}$ in SCM is
\begin{equation}
\theta_{ij}=r_i+r_j-R_{\mathrm{SCM}}.
\end{equation}
The SCM Hamiltonian is then
\begin{equation}
\begin{split}
H_{\mathrm{SCM}}(G)=&\sum_{i<j}\theta_{ij}a_{ij}=\sum_{i<j}(r_i+r_j-R_{\mathrm{SCM}})a_{ij}\\ =&\sum_{i=1}^N k_ir_i - MR_{\mathrm{SCM}},
\end{split}
\end{equation}
where $k_i$ is the degree of node $i$ and $M$ is the total number of edges in the graph $G$.

\begin{figure}
	\centerline{\includegraphics[width=85mm]{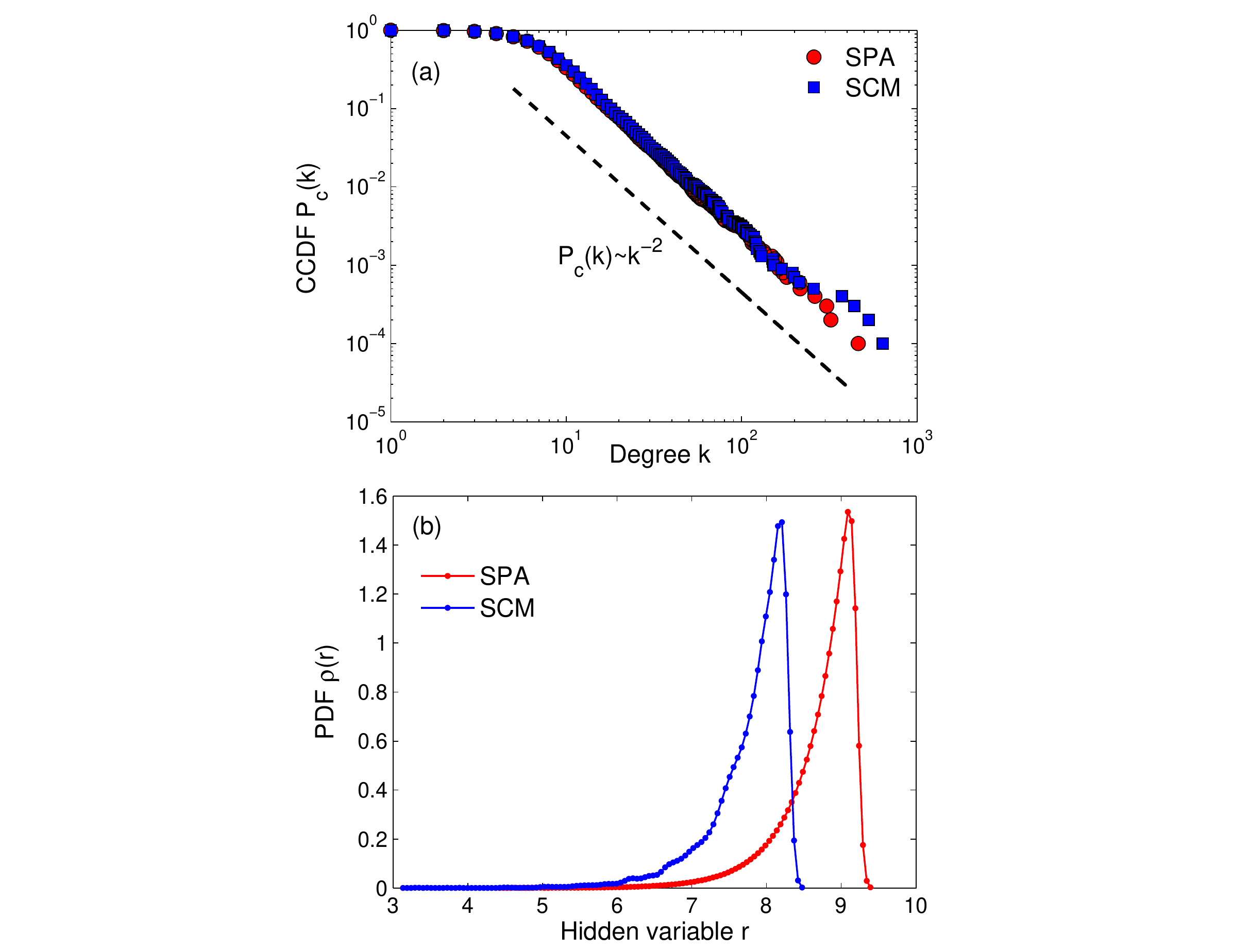}}
	\caption{\small \textbf{SPA and SCM networks.} Panel~(a) shows the empirical complementary cumulative degree distribution functions (CCDF) $P_c(k)=\sum_{k'>k}\mathbb{P}(k')$ for two networks of size $N=10^4$ generated by SPA and SCM with $\bar{k}=10$ and $\gamma=3$, and the corresponding power-law fit. As expected, $P_c(k)\sim k^{-\gamma+1}$. Panel~(b) shows the empirical probability density functions (PDF) of the hidden variables $r_i$ in these two networks.}
	\label{Fig1}
\end{figure}

Figure~\ref{Fig1}(a) shows the degree distribution in a network generated by SCM, which is identical to the degree distribution in an SPA network generated with the same parameters. As expected, both are power-laws with exponent $\gamma$. Figure~\ref{Fig1}(b) shows the empirical distributions of the hidden variables $\rho_{\mathrm{SPA}}(r)$ and $\rho_{\mathrm{SCM}}(r)$ in the generated networks. Although both distributions are highly skewed to the right, there is a clear discrepancy between them. In the next section we fix this discrepancy by introducing a \textit{shifted} SCM model, which is strongly equivalent to SCM, but has the same distribution of hidden variables as SPA.

\subsection{Shifted Soft Configuration Model}\label{sec:SCM+}

While the hidden variables in SCM are random, in SPA they are deterministic. Nevertheless we can readily overcome this technical obstruction that hinders the comparison of hidden variable distributions in the two models.

Let $r_*(N)$ denote one of the hidden variables $r_1(N),\ldots,r_N(N)$ in SPA chosen uniformly at random at time $N\gg1$. The CDF of $r_*(N)$ is then
\begin{equation}
\begin{split}
F_{r_*(N)}(r)=&\mathbb{P}(r_*(N)\leq r)=\frac{|\{i : r_i(N)\leq r\}|}{N}\\ =&\frac{|\{i : \beta\ln i+(1-\beta)\ln N\leq r\}|}{N}\\
=&\frac{|\{i : i\leq e^{\frac{r}{\beta}-\frac{1-\beta}{\beta}\ln N}\}|}{N} =\frac{1}{N}\hspace{-14mm}\sum_{\hspace{14mm}i\leq e^{\frac{r}{\beta}-\frac{1-\beta}{\beta}\ln N}}\hspace{-14mm}1\\ \approx& \frac{1}{N}\hspace{-12mm}\int\limits_{0}^{\hspace{12mm}e^{\frac{r}{\beta}-\frac{1-\beta}{\beta}\ln N}}\hspace{-12mm}1di =e^{\frac{r-\ln N}{\beta}}.
\end{split}
\end{equation}
When $N$ is large, the SPA hidden variables $r_1(N),\ldots,r_N(N)$ can be viewed as being  approximately i.i.d. samples from $F_{r_*(N)}(r)$. This distribution has the following PDF
\begin{equation}\label{rhoPA}
\rho_{\mathrm{SPA}}(r)=\frac{d}{dr}F_{r_*(N)}(r)=\frac{1}{\beta}e^{\frac{r-\ln N}{\beta}},
\end{equation}
which is structurally similar to the  distribution of hidden variables in SCM
\begin{eqnarray}
\rho_{\mathrm{SCM}}(r)&\approx&\alpha e^{\alpha(r-R_{\mathrm{SCM}})}=\frac{1}{\beta}e^{\frac{r-\ln N+\sigma}{\beta}}\text{, where}\\
\sigma&=&\ln\bar{k}(1-\beta)^2.
\end{eqnarray}

It is readily verifiable that the approximate supports of   $\rho_{SPA}(r)$ and $\rho_{SCM}(r)$, \ie segments that contain  almost all probability mass of these distributions, are
\begin{align}
\mbox{supp}\;\rho_{PA}&=[(1-\beta)\ln N, \ln N], \label{supprhoPA}\\
\mbox{supp}\;\rho_{SCM}&=[(1-\beta)\ln N-\sigma,\ln N-\sigma].\label{supp scm}
\end{align}
Therefore, it immediately follows from (\ref{rhoPA})-(\ref{supp scm}) that $\rho_{SPA}(r)$ is obtained from  $\rho_{SCM}(r)$ by translation by $\sigma$. This motivates the shifted SCM model, denoted SCM$^+$, which generates networks of size $N$ with average degree $\bar{k}$ and power-law exponent $\gamma > 2$, as follows:
\begin{enumerate}
	\item For $i=1,\ldots,N$, assign to node $i$ hidden variable $r_i^+$ by, first, sampling $r_i\sim\rho_{\mathrm{SCM}}(r)$, and then shifting $r_i^+=r_i+\sigma$, where $\sigma=\ln\bar{k}(1-\beta)^2$.
	\item Connect nodes $i$ and $j$, $j\neq i$, with probability
	\begin{equation}\label{eq:scm+ p_ij}
	p_{ij}=\frac{1}{1+e^{r_i^++r_j^+-R_{\mathrm{SCM}^+}}},
	\end{equation}
	where $R_{\mathrm{SCM}^+}$ is given by~(\ref{eq:RSCM+}).
\end{enumerate}

By construction, the distributions of hidden variables in SCM$^+$ and SPA are identical
\begin{equation}
\rho_{\mathrm{SCM}^+}(r)=\rho_{\mathrm{SCM}}(r-\sigma)=\rho_{\mathrm{SPA}}(r).
\end{equation}
To make SCM$^+$ equivalent to SCM, we need to choose $R_{\mathrm{SCM}^+}$ appropriately. If two nodes have hidden variables $r$ and $\acute{r}$ in SCM, then in SCM$^+$ the values of these hidden variables are $r^+=r+\sigma$ and  $\acute{r}^+=\acute{r}+\sigma$. The two models will be strongly equivalent, \ie will generate graphs $G\in\mathcal{G}_N$ with equal probabilities $\mathbb{P}_{\mathrm{SCM}}(G)=\mathbb{P}_{\mathrm{SCM}^+}(G)$, if the connection probabilities $p_{\mathrm{SCM}}(r,\acute{r})$  and $p_{\mathrm{SCM}^+}(r^+,\acute{r}^+)$ are the same. This leads to the following equation for $R_{\mathrm{SCM}^+}$
\begin{equation}
r+\acute{r}-R_{\mathrm{SCM}}=r^++\acute{r}^+-R_{\mathrm{SCM}^+}.
\end{equation}
Therefore,
\begin{equation}
R_{\mathrm{SCM}^+}=\ln N + \sigma=
\ln N\bar{k}(1-\beta)^2\label{eq:RSCM+}.
\end{equation}

It is convenient to work with SPA and SCM$^+$ (instead of SCM) since not only the degree distributions in the networks generated by these two models match, but also the distributions of hidden variables are the same. Our next step is to adjust SPA so that it becomes strongly equivalent to SCM$^+$ (and, therefore, to SCM).

\subsection{Bridging SPA and SCM$^+$}\label{sec:tildeSPA}

Matching degree distributions is a necessary but, of course, not sufficient condition for model equivalence. In SPA, a link between nodes $i$ and $j<i$ may appear only at time $i$ upon the birth of the younger node. We refer to such links --- appearing at time $i$ and connecting new node $i$ to already existent nodes --- as ``external'' links. We make the SPA model equivalent to SCM$^+$ by also allowing ``internal'' links that appear at time $i$ and connect old nodes $a$ and $b$, where $a,b<i$.
Namely, we define the model $\widetilde{\mathrm{SPA}}$ that generates growing networks up to some size $N$ with average degree $\bar{k}$ and power-law exponent $\gamma > 2$, as follows. For each new node $i=1,\ldots,N$:
\begin{enumerate}
\item Assign to node $i$ hidden variable $r_i=\ln i$.
\item Update the values of hidden variables of all existing nodes $j<i$ by setting $r_j(i)=\beta r_j+(1-\beta)r_i$, where $\beta=\frac{1}{\gamma-1}$.
\item Connect node $i$ to each existing node $j<i$ with probability
\begin{equation}
p_{ij}^{\mathrm{ext}}=\frac{1}{1+e^{r_j(i)+r_i-R_i^{\mathrm{ext}}}},
\end{equation}
where
\begin{equation}
R_i^{\mathrm{ext}}=r_i-\ln\frac{1-e^{-(1-\beta)r_i}}{m_{\mathrm{ext}}(1-\beta)},
\end{equation}
and $m_{\mathrm{ext}}$ given by~(\ref{eq:mext}).
\item Connect each pair of existing nodes $a,b<i$ with probability
\begin{equation}
p_{ab}^{\mathrm{int}}(i)=\frac{1}{1+e^{r_a(i)+r_b(i)-R^{\mathrm{int}}}},
\end{equation}
where
\begin{equation}
R^{\mathrm{int}}=\ln{m_{\mathrm{int}}(1-\beta)},
\end{equation}
and $m_{\mathrm{int}}$ given by~(\ref{eq:mint}).
\end{enumerate}
In Step~4, we scan all pairs of existing nodes and attempt to connect even those nodes which are already connected.  The $\widetilde{\mathrm{SPA}}$ model thus allows multi-edges. In large sparse ($\bar{k} \ll N$) networks, however, the proportion of multi-edges is small. For example, the expected ratio of multi-edges in  $\widetilde{\mathrm{SPA}}$ networks of size $N=10^2, 10^3,$ and $10^4$ with $\bar{k}=10$ and $\gamma=2.5$ is, respectively, $7\%$, $4\%$ and $2\%$. We can therefore ignore the multi-edge effect.
The choices for $m_{\mathrm{ext}}$ and $m_{\mathrm{int}}$ (Eqs.~(\ref{eq:mext}) and (\ref{eq:mint})) are explained below.

First, a necessary (but not sufficient) condition for the equivalence of two models is that the expected minimum degrees in the two models must be the same. The expected minimum degree in large networks generated by SCM (and therefore by SCM$^+$) is $\bar{k}(1-\beta)$ \cite{KrPaKi10}. Thus, we have the following condition
\begin{equation}\label{eq:cond1}
\langle k_{\min}\rangle_{\widetilde{\mathrm{SPA}}}=\bar{k}(1-\beta).
\end{equation}
Let us now compute the expected degree of a new node $i$ upon its birth in $\widetilde{\mathrm{SPA}}$. For large $i$, $R_i^{\mathrm{ext}}\approx r_i+\ln m_{\mathrm{ext}}(1-\beta)$, and using the classical limit for the Fermi-Dirac distribution, we get
\begin{equation}\label{eq:<k>birth}
\begin{split}
\kappa_i(i)=&\sum_{j<i}p_{ij}^{\mathrm{ext}}\approx\int_0^i\frac{dj}{1+e^{r_j(i)+r_i-R_i^{\mathrm{ext}}}}\\
\approx&  \int_0^i\frac{dj}{1+e^{\beta r_j+(1-\beta)r_i-\ln m_{\mathrm{ext}}(1-\beta)}}\\
=&\int_0^i\frac{dj}{1+\frac{j^{\beta}i^{1-\beta}}{m_{\mathrm{ext}}(1-\beta)}}\approx \frac{m_{\mathrm{ext}}(1-\beta)}{i^{1-\beta}}\int_0^ij^{-\beta}dj\\ =&m_{\mathrm{ext}}.
\end{split}
\end{equation}
Every new node thus establishes on average $m_{\mathrm{ext}}$ links, and, as time goes, its degree may only increase. This means that $m_{\mathrm{ext}}$ is the expected minimum degree in $\widetilde{\mathrm{SPA}}$, $\langle k_{\min}\rangle_{\widetilde{\mathrm{SPA}}}=m_{\mathrm{ext}}$, and therefore
\begin{equation}\label{eq:mext}
m_{\mathrm{ext}}=\bar{k}(1-\beta).
\end{equation}

\begin{figure}
	\centerline{\includegraphics[width=85mm]{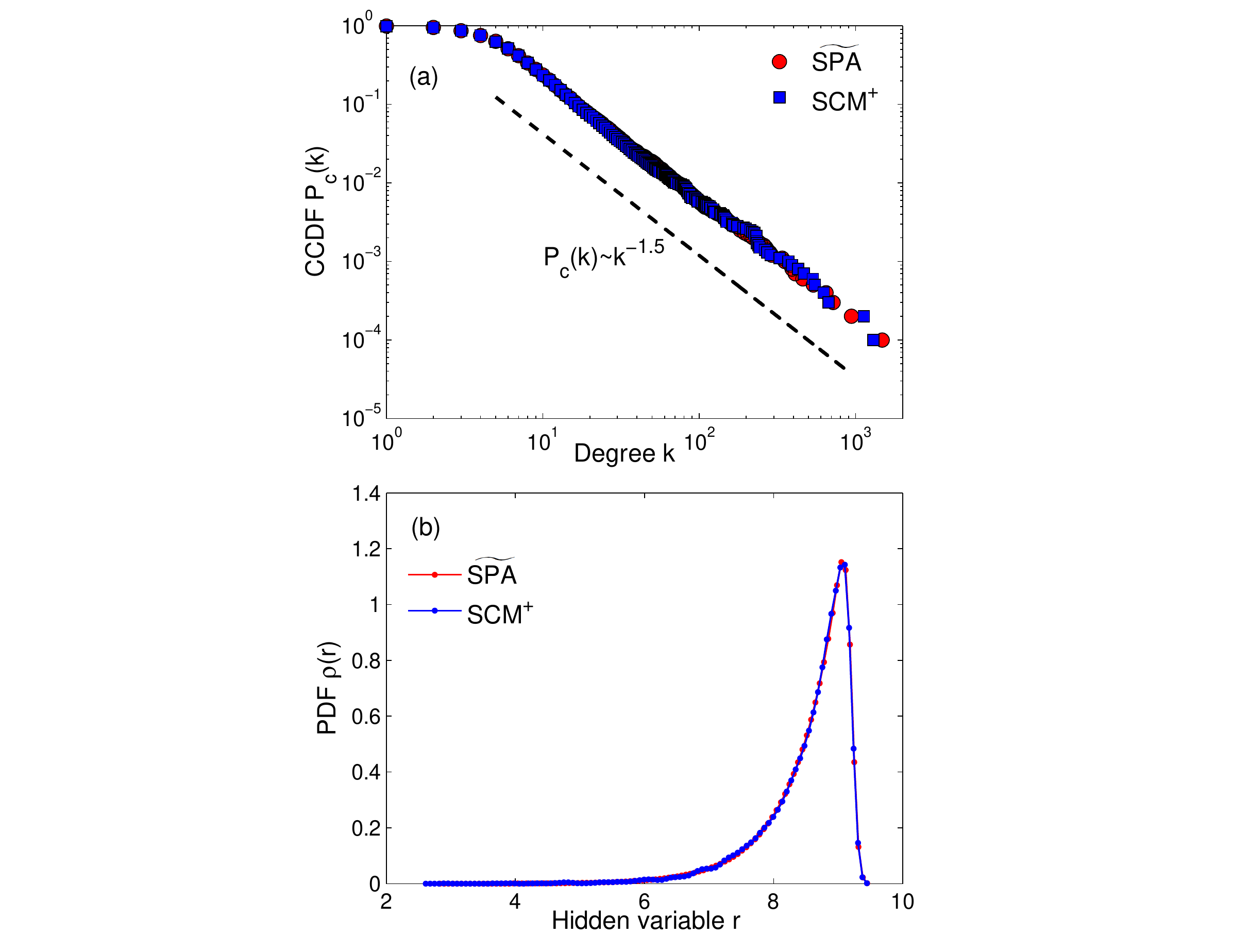}}
	\caption{\small \textbf{$\widetilde{\bf{SPA}}$ and SCM$^+$ networks.} Panel~(a) shows the empirical CCDF $P_c(k)=\sum_{k'>k}\mathbb{P}(k')$ for two networks of size $N=10^4$ generated by $\widetilde{\mathrm{SPA}}$ and SCM$^+$ with $\bar{k}=10$ and $\gamma=2.5$, and the corresponding power-law fit. As expected, for both networks $P_c(k)\sim k^{-\gamma+1}$. Panel~(b) shows the perfect match between the PDFs of the hidden variables in these two networks, illustrating the theoretical result $\rho_{\widetilde{\mathrm{SPA}}}(r)=\rho_{\mathrm{SCM}^+}(r)$.}
	\label{Fig2}
\end{figure}

Another necessary condition (that helps to determine $m_{\mathrm{int}}$) for the equivalence between $\widetilde{\mathrm{SPA}}$  and SCM$^+$ is that the expected average degrees in both models must be the same.  That is, if in SCM$^+$ the expected average degree $\langle \bar{k}\rangle_{\mathrm{SCM}^+}$ equals $\bar{k}$, then we must have
\begin{equation}\label{eq:cond2}
\langle \bar{k}\rangle_{\widetilde{\mathrm{SPA}}}=\bar{k}.
\end{equation}
Let $\bar{L}^{\mathrm{int}}_i$ denote the expected number of internal links generated at time $i$. Then, the expected total number of links generated at time $i$ is $m_{\mathrm{ext}}+\bar{L}^{\mathrm{int}}_i$, and the expected average degree in the network is given by
\begin{equation}\label{eq:expavedeg}
\langle \bar{k}\rangle_{\widetilde{\mathrm{SPA}}}\approx\frac{2}{N}\int_0^N\left(m_{\mathrm{ext}}+\bar{L}^{\mathrm{int}}_i\right)di.
\end{equation}
For large $i$,
\begin{equation}\label{eq:Lint}
\begin{split}
\bar{L}^{\mathrm{int}}_i=& \sum_{a<i}\sum_{b<a}p_{ab}^{\mathrm{int}}(i)
\approx\int_0^i\int_0^a\frac{dbda}{1+e^{r_a(i)+r_b(i)-R^{\mathrm{int}}}}\\
=&\int_0^i\int_0^a\frac{dbda}{1+e^{\beta r_a + \beta r_b + 2(1-\beta)r_i-\ln m_{\mathrm{int}}(1-\beta)}}\\
=& \int_0^i\int_0^a\frac{dbda}{1+\frac{a^{\beta}b^{\beta}i^{2(1-\beta)}}{m_{\mathrm{int}}(1-\beta)}}\\ \approx& \frac{m_{\mathrm{int}}(1-\beta)}{i^{2(1-\beta)}}\int_0^i\int_0^a a^{-\beta}b^{-\beta}dbda\\
=&\frac{m_{\mathrm{int}}}{2(1-\beta)}.
\end{split}
\end{equation}
Combining (\ref{eq:mext})-(\ref{eq:Lint}), we get $m_{\mathrm{int}}$
\begin{equation}\label{eq:mint}
m_{\mathrm{int}}=\bar{k}(1-\beta)(2\beta-1),
\end{equation}
which is positive if $\beta\in(1/2,1)$, or, equivalently, $\gamma\in(2,3)$. We note that $2<\gamma<3$ is exactly the range of power-law exponents empirically observed in most real networks~\cite{AlBa02}. Figure~\ref{Fig2} shows the perfect match between the distributions of node degrees and hidden variables in $\widetilde{\mathrm{SPA}}$ and SCM$^+$ networks with $\gamma=2.5$. Recall that the match between the hidden variable distributions is the first condition~(\ref{eq:same-rho}) for two ensembles of random graphs with hidden variables to be equivalent.

We note that as $\beta\rightarrow1/2$, or, equivalently, $\gamma\rightarrow3$, $\widetilde{\mathrm{SPA}}$ becomes manifestly identical to SPA. Indeed, in this case,
\begin{equation}
m_{\mathrm{ext}}\rightarrow m=\frac{\bar{k}}{2} \hspace{5mm}
\mbox{and} \hspace{5mm}m_{\mathrm{int}}\rightarrow 0,
\end{equation}
which means
\begin{equation}
R^{\mathrm{int}}\rightarrow -\infty \hspace{5mm}
\mbox{and} \hspace{5mm} p_{ab}^{\mathrm{int}}(i)\rightarrow 0,~\forall i.
\end{equation}
The other limiting case $\beta\rightarrow1$ ($\gamma\rightarrow2$) is analyzed in Appendix~\ref{A1}.

It is important to realize that by choosing $m_{\mathrm{ext}}$ and $m_{\mathrm{int}}$ according to (\ref{eq:mext}) and (\ref{eq:mint}) we only satisfied two necessary conditions, but we did not actually prove that $\widetilde{\mathrm{SPA}}$ is equivalent to SCM$^+$. We prove  this in the next section.

\subsection{$\widetilde{\bf{SPA}}$ and SCM$^+$ are strongly equivalent}
\label{sec:tSPA=SCM+}

Since the distributions of hidden variables in $\widetilde{\mathrm{SPA}}$ and SCM$^+$ are the same, \ie $\rho_{\widetilde{\mathrm{SPA}}}(r)=\rho_{\mathrm{SCM}^+}(r)$, to prove the strong equivalence between the two models, we need to show that the connection probabilities in $\widetilde{\mathrm{SPA}}$ and SCM$^+$ are also the same. More precisely, if at time $N$ the values of hidden variables of nodes $j$ and $i>j$ in $\widetilde{\mathrm{SPA}}$ are $r$ and $\acute{r}>r$, then  the probability that these two nodes are connected $p_{\widetilde{\mathrm{SPA}}}(i,j)$ must be equal to the connection probability $p_{\mathrm{SCM}^+}(r,\acute{r})$ of nodes with hidden variables $r$ and $\acute{r}$ in SCM$^+$. In what follows, we compute these probabilities in large sparse graphs ($N\gg1$, $\bar{k} \ll N$) and show that they indeed coincide. Throughout this section we assume that $\beta\in(1/2,1)$ ($\gamma\in(2,3)$).

Let $p^*_{\widetilde{\mathrm{SPA}}}(i,j)$ denote the probability that nodes $i$ and $j$ are not connected in $\widetilde{\mathrm{SPA}}$, then
\begin{equation}\label{eq:p_paintetx}
\begin{split}
&p_{\widetilde{\mathrm{SPA}}}(i,j)=1-p^*_{\widetilde{\mathrm{SPA}}}(i,j)\\ &=1-(1-p_{ij}^{\mathrm{ext}})(1-p_{ij}^{\mathrm{int}}(i+1))\ldots(1-p_{ij}^{\mathrm{int}}(N))\\
&\approx p_{ij}^{\mathrm{ext}} + \sum_{s>i}^N p_{ij}^{\mathrm{int}}(s)\approx p_{ij}^{\mathrm{ext}} +\int_i^N p_{ij}^{\mathrm{int}}(s)ds.
\end{split}
\end{equation}
Let us compute the integral first
\begin{equation}
\label{eq:integral}
\begin{split}
\int_i^N &p_{ij}^{\mathrm{int}}(s)ds=\int_i^N
\frac{ds}{1+e^{r_i(s)+r_j(s)-R^{\mathrm{int}}}}\\
=& \int_i^N
\frac{ds}{1+e^{\beta r_i+\beta r_j +2(1-\beta)\ln s-\ln m_{\mathrm{int}}(1-\beta)}}.
\end{split}
\end{equation}
Since $r$ and $\acute{r}$ are the values of hidden variables of nodes $j$ and $i$ at time $N$, we have
\begin{align}\label{eq:randr'}
r=&r_j(N)=\beta r_j+(1-\beta)\ln N,\\
\acute{r}=&r_i(N)=\beta r_i+(1-\beta)\ln N. \label{eq:ri(N)}
\end{align}
Finding from these equations $r_i$ and $r_j$ and substituting them into (\ref{eq:integral}), we obtain
\begin{equation}\label{eq:integral2}
\int_i^N p_{ij}^{\mathrm{int}}(s)ds=\int_i^N\frac{ds}{1+\frac{e^{r+\acute{r}}}{m_{\mathrm{int}}(1-\beta)}\left(\frac{s}{N}\right)^{2(1-\beta)}}.
\end{equation}
To proceed with analytic approximation, we need to use the classical limit for the Fermi-Dirac distribution, \ie to drop the term $1$ in the denominator. We have already used this approximation in (\ref{eq:<k>birth}) and (\ref{eq:Lint}). In those cases, the second terms in the denominator were fully deterministic, and it was readily verifiable that they are much larger than $1$, and therefore the approximations hold. In (\ref{eq:integral2}) however, both $r$ and $\acute{r}$ are random, $r,\acute{r}\sim\rho_{\mathrm{SCM}^+}$, and a certain caution is required.

In what follows, we show that the expected value of $e^r$ in SCM$^+$ scales as $N$, and, therefore, $\frac{e^{r+\acute{r}}}{N^{2(1-\beta)}}\propto N^{2\beta}\geq N\gg1$. Indeed,
\begin{equation}\label{eq:<>}
\begin{split}
\langle e^r\rangle=&\int_0^{R_{\mathrm{SCM}}}e^{x+\sigma}\rho_{\mathrm{SCM}}(x)dx\\
\approx &\frac{e^{\sigma}}{\beta}\int_0^{R_{\mathrm{SCM}}}e^x e^{\frac{x-R_{\mathrm{SCM}}}{\beta}}dx\\
=&\frac{\bar{k}(1-\beta)^2}{\beta e^{\frac{R_{\mathrm{SCM}}}{\beta}}}\int_0^{R_{\mathrm{SCM}}}e^{(1+\frac{1}{\beta})x}dx\\ =&\frac{\bar{k}(1-\beta)^2}{(1+\beta) e^{\frac{R_{\mathrm{SCM}}}{\beta}}}\left(e^{(1+\frac{1}{\beta})R_{\mathrm{SCM}}}-1\right)\\
\approx& \frac{\bar{k}(1-\beta)^2}{1+\beta}e^{R_{SCM}}=\frac{N}{1+\beta}\propto N.
\end{split}
\end{equation}
Thus,
\begin{equation}\label{eq:integral3}
\begin{split}
\int_i^N &p_{ij}^{\mathrm{int}}(s)ds\approx\frac{(1-\beta)m_{\mathrm{int}}}{e^{r+\acute{r}}N^{-2(1-\beta)}}\int_i^N s^{-2(1-\beta)}ds\\
=&\frac{(1-\beta)m_{\mathrm{int}}}{(2\beta-1)e^{r+\acute{r}}N^{-2(1-\beta)}}\left(N^{2\beta-1}-i^{2\beta-1}\right)\\
=&\frac{\bar{k}(1-\beta)^2N}{e^{r+\acute{r}}}-\frac{\bar{k}(1-\beta)^2N^{2(1-\beta)}i^{2\beta-1}}{e^{r+\acute{r}}}.
\end{split}
\end{equation}

The probability of the external link is
\begin{equation}\label{eq:pext}
\begin{split}
p_{ij}^{\mathrm{ext}}=&\frac{1}{1+e^{r_j(i)+r_i-R_i^{\mathrm{ext}}}}\\
\approx&\frac{1}{1+e^{\beta r_j+(1-\beta)r_i-\ln m_{\mathrm{ext}}(1-\beta)}}\\
=&\frac{1}{1+\frac{e^{r+\frac{1-\beta}{\beta}\acute{r}}}{m_{\mathrm{ext}}(1-\beta)N^{\frac{1-\beta}{\beta}}}}\approx\frac{\bar{k}(1-\beta)^2N^{\frac{1-\beta}{\beta}}}{e^{r+\frac{1-\beta}{\beta}\acute{r}}},
\end{split}
\end{equation}
where the last approximation holds because, using (\ref{eq:<>}), $\left.{e^{r+\frac{1-\beta}{\beta}\acute{r}}}\right/{N^{\frac{1-\beta}{\beta}}}\propto N$. Combining (\ref{eq:p_paintetx}), (\ref{eq:integral3}) and (\ref{eq:pext}), we obtain that the probability that nodes $i$ and $j$ in $\widetilde{\mathrm{SPA}}$ are connected is
	\begin{multline}\label{eq:prepPA}
	p_{\widetilde{\mathrm{SPA}}}(i,j)=\frac{\bar{k}(1-\beta)^2N}{e^{r+\acute{r}}}\\ -\frac{\bar{k}(1-\beta)^2N^{2(1-\beta)}i^{2\beta-1}}{e^{r+\acute{r}}}+\frac{\bar{k}(1-\beta)^2N^{\frac{1-\beta}{\beta}}}{e^{r+\frac{1-\beta}{\beta}\acute{r}}}.
	\end{multline}

To compare this probability with the connection probability in SCM$^+$, we need to rewrite it fully in terms of hidden variables $r$ and $\acute{r}$. From (\ref{eq:ri(N)}), $r_i=\frac{\acute{r}}{\beta}-\frac{1-\beta}{\beta}\ln N$. Therefore,
\begin{equation}
i=e^{\frac{\acute{r}}{\beta}-\frac{1-\beta}{\beta}\ln N}=e^{\frac{\acute{r}}{\beta}}N^{-\frac{1-\beta}{\beta}}.
\end{equation}
Substituting this expression into (\ref{eq:prepPA}), we obtain after some algebra that the last two terms cancel out, and
\begin{equation}\label{eq:paextintfinal}
p_{\widetilde{\mathrm{SPA}}}(i,j)=\frac{\bar{k}(1-\beta)^2N}{e^{r+\acute{r}}}.
\end{equation}

\begin{figure}
	\centerline{\includegraphics[width=85mm]{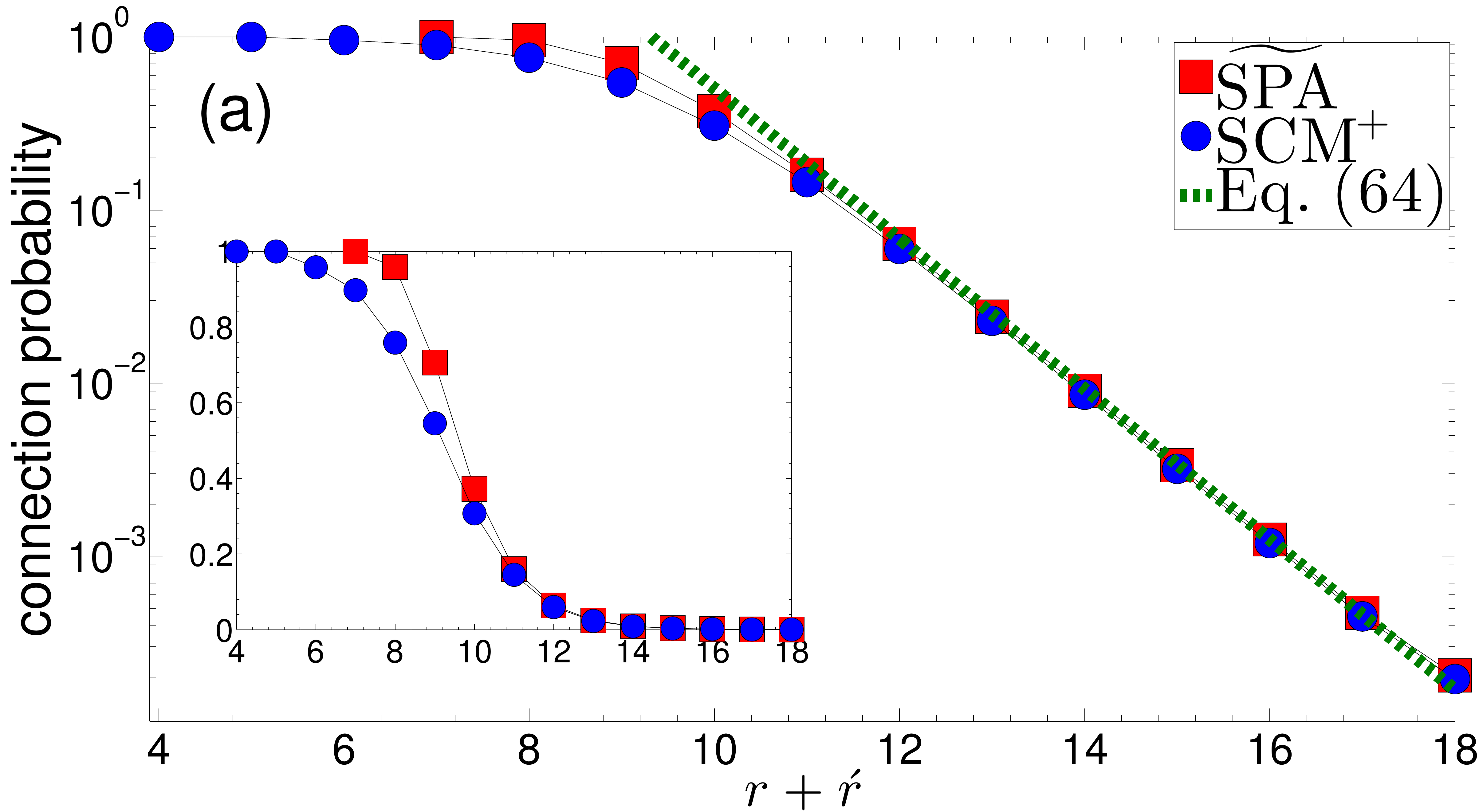}}
	\centerline{\includegraphics[width=85mm]{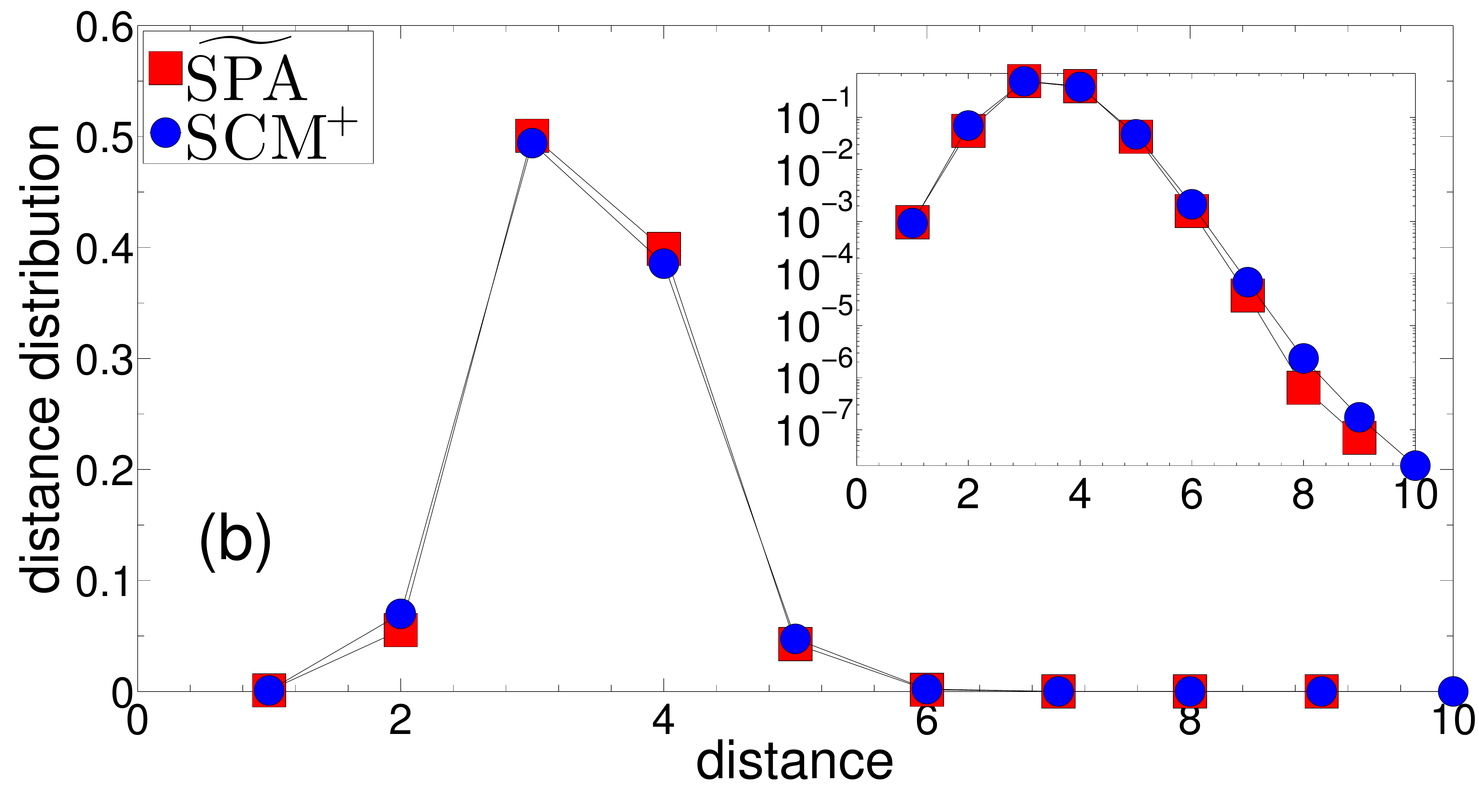}}
	\caption{\small \textbf{Connection probabilities  and vertex-to-vertex distance distribution in $\widetilde{\bf{SPA}}$ and SCM$^+$ networks.} Panel (a) shows the empirical connection probabilities in networks of size $N=10^4$ generated by $\widetilde{\mathrm{SPA}}$ and SCM$^+$ with $\bar{k}=10$ and $\gamma=2.5$. The $y$-axis in the main plot is in logarithmic scale. As expected, the connection probabilities match remarkably well. The plot also shows the corresponding approximation given in (\ref{eq:p_scm}). As expected, this approximation holds very well for sufficiently large values of $r+\acute{r}$, which correspond to the vast majority of node pairs, cf.~Fig.~\ref{Fig2}(b) and Appendix~\ref{A2}.  Panel (b) shows the distance distributions in the same networks. The $y$-axis in the inset is in logarithmic scale. As a consequence of the equivalence of $\widetilde{\mathrm{SPA}}$ and SCM$^+$, the distance distributions also match remarkably well, as expected. In all cases the results were averaged across $100$ networks.}
	\label{c_prob}
\end{figure}

It remains to show that (\ref{eq:paextintfinal}) is, in fact, the connection probability in SCM$^+$. Indeed,
\begin{equation}\label{eq:p_scm}
\begin{split}
p_{\mathrm{SCM}^+}(r,\acute{r})&=\frac{1}{1+e^{r+\acute{r}-R_{\mathrm{SCM}^+}}}\\ =&\frac{1}{1+e^{r+\acute{r}-\ln N\bar{k}(1-\beta)^2}}\\
&=\frac{1}{1+\frac{e^{r+\acute{r}}}{N\bar{k}(1-\beta)^2}}\approx\frac{\bar{k}(1-\beta)^2N}{e^{r+\acute{r}}},
\end{split}
\end{equation}
where the last approximation holds because $\frac{e^{r+\acute{r}}}{N}\propto N$. In Appendix~\ref{A2}, we show the high accuracy of this approximation with simulation. Fig.~\ref{c_prob}(a) juxtaposes the empirical connection probabilities in $\widetilde{\mathrm{SPA}}$ and SCM$^+$ networks, and the corresponding approximation in~(\ref{eq:p_scm}).
Recall that the match between the connection probabilities is the second condition~(\ref{eq:same-p}) for two ensembles of random graphs with hidden variables to be equivalent.

Thus, we proved that, for large networks, $\widetilde{\mathrm{SPA}}$ is equivalent to SCM$^+$ in the strong sense. That is, for any $N\gg1$,  $\widetilde{\mathrm{SPA}}$ and SCM$^+$ generate graphs $G\in\mathcal{G}_N$ with the same probability,
$
\mathbb{P}_{\widetilde{\mathrm{SPA}}}(G)=\mathbb{P}_{\mathrm{SCM}^+}(G).
$
Therefore, the expected values of \emph{all} graph properties in the ensemble, not only of the degree distributions in Fig.~\ref{Fig2}(a), are the same. As an example, Fig.~\ref{c_prob}(b) shows the vertex-to-vertex distance distribution $d(l)$, which is the distribution of hop lengths $l$ of shortest paths between nodes in the network, or the probability that a random pair of nodes are at the distance of $l$ hops from each other.

The following diagram summarizes the relationships between the four considered network models
\begin{equation}
\begin{array}[c]{ccccccc}
\mathrm{SPA}&\rightsquigarrow&\widetilde{\mathrm{SPA}}&\stackrel{s}{\approx}&\mathrm{SCM}^+&\stackrel{s}{=}&\mathrm{SCM},
\end{array}
\end{equation}
where $\stackrel{s}{=}$ denotes the strong model equivalence, $\stackrel{s}{\approx}$ is an approximate strong equivalence that becomes exact in the sparse graph limit ($N\rightarrow\infty$, $\bar{k} \ll N$),
and $\rightsquigarrow$ denotes the model transformation allowing internal links. Furthermore,
\begin{equation}
\mathrm{SPA}\bigg|_{\beta=1/2}\stackrel{s}{=}\widetilde{\mathrm{SPA}}\bigg|_{\beta=1/2},
\end{equation}
that is, SPA and $\widetilde{\mathrm{SPA}}$ are strongly equivalent if $\beta=1/2$, i.e., $\gamma=3$.

\subsection{Hamiltonians of SCM$^+$, $\widetilde{\bf{SPA}}$, and SPA}\label{sec:all-erg-hamiltonians}

As discussed in Section~\ref{sec:SCM}, SCM is the ERGM model with Hamiltonian
\begin{equation}
\begin{split}
H_{\mathrm{SCM}}=&\sum_{i=1}^N k_ir_i-MR_{\mathrm{SCM}}\\
=&\sum_i^N k_ir_i-\frac{(\ln N - \sigma)}{2}\sum_{i=1}^Nk_i.
\end{split}
\end{equation}
Since SCM$^+$ is strongly equivalent to SCM --- the two models differ only by parametrization of hidden variables --- SCM$^+$ must have the same Hamiltonian. Indeed,
\begin{equation}
\begin{split}
H_{\mathrm{SCM}^+}=&\sum_{i=1}^N k_ir_i^+-MR_{\mathrm{SCM}^+}\\
=&\sum_{i=1}^N k_i(r_i+\sigma)-\frac{(\ln N + \sigma)}{2}\sum_{i=1}^Nk_i\\
=&\sum_{i=1}^N k_ir_i-\frac{(\ln N - \sigma)}{2}\sum_{i=1}^Nk_i=H_{\mathrm{SCM}}.
\end{split}
\end{equation}
Further, since $\widetilde{\mathrm{SPA}}$ is strongly equivalent to SCM$^+$ and the distributions of hidden variables in these two models are the same, the ERGM Hamiltonian of $\widetilde{\mathrm{SPA}}$ is
\begin{equation}\label{eq:HPAextint}
\begin{split}
H_{\widetilde{\mathrm{SPA}}}=&
\sum_{i=1}^N k_ir_i-MR_{\mathrm{SCM}^+}\\
=&\sum_{i=1}^N k_ir_i-\frac{(\ln N + \sigma)}{2}\sum_{i=1}^Nk_i\\
=&\sum_{i=1}^N k_ir_i-\frac{\ln \left(N\bar{k}(1-\beta)^2\right)}{2}\sum_{i=1}^Nk_i.
\end{split}
\end{equation}

Finally, since $\widetilde{\mathrm{SPA}}$ becomes manifestly identical to SPA as $\beta\rightarrow1/2$, or, equivalently, as $\gamma\rightarrow3$, we can write the ERGM Hamiltonian for SPA
\begin{equation}\label{eq:HPA}
H_{\mathrm{SPA}}=\sum_{i=1}^N k_ir_i-\frac{\ln\left( N\bar{k}/4\right)}{2}\sum_{i=1}^Nk_i.
\end{equation}
This result means that if $\beta=1/2$ ($\gamma=3$), then, in the sparse graph limit,
SPA is exactly ERGM with Hamiltonian (\ref{eq:HPA}). We note that this case corresponds to the original
Barab\'{a}si--Albert model with the scaling exponent $\gamma=3$ \cite{BarAlb99}. If $\beta\neq1/2$, then (\ref{eq:HPA}) is an approximate Hamiltonian of SPA. To the best of our knowledge, this is the first result where the preferential attachment model is represented as an exponential random graph model with explicitly written Hamiltonian. Even more remarkably, as we show in the next section, a Hamiltonian that is very similar to the ERGM Hamiltonian (\ref{eq:HPA}) describes the Hamiltonian dynamics of growing networks in SPA.

\section{Hamiltonian Dynamics of SPA}
\label{sec:HamDynSPA}

The key idea in deriving the ERGM Hamiltonian of SPA (\ref{eq:HPA}) was to construct a modified model $\widetilde{\mathrm{SPA}}$ that: a) is strongly equivalent to a graph ensemble with a known Hamiltonian; and b) coincides with SPA under certain values of the model parameters. Here we adopt a similar strategy: we study the Hamiltonian dynamics of growing $\widetilde{\mathrm{SPA}}$-networks, and the corresponding results for the Hamiltonian dynamics of SPA are obtained as a special case with $\beta=1/2$.

The ERGM Hamiltonian of $\widetilde{\mathrm{SPA}}$ (\ref{eq:HPAextint}) suggests that the canonical coordinates $\{q_i,p_i\}$ of a growing network are the node degrees $k_i$ and hidden variables $r_i$. An immediate technical problem we face, however, is that both node degrees $k_i$ and network time $i$ are discrete. We overcome this obstruction as follows. First,  inspired by the mapping between the hyperbolic and de Sitter spaces in~\cite{KrKi12}, we define
\begin{equation}\label{eq:rescost}
t=\beta\ln i,
\end{equation}
and treat $t$ as a continuous time. The geometric duality between de Sitter spacetime, which is asymptotically the spacetime of our accelerating universe, and hyperbolic space, which is a latent space underlying real complex networks \cite{KrPaKi10,PKSBK12} allows to interpret $t$ as the \textit{rescaled cosmological time}. Second, instead of the exact (discrete) degree $k_i(t)$ of node $i$, born at time $t_i=\beta\ln i$, at a later time $t>t_i$, we use its \textit{expected} degree $\kappa_i(t)$, which is a continuous function of $t$, for $t>t_i$.

Our next goal is to derive the time evolution of the canonical coordinates $\{\kappa_i(t),r_i(t)\}$ in the growing SPA- and $\widetilde{\mathrm{SPA}}$-networks. Given that in network time $i$, $r_j(i)=\beta r_j+(1-\beta)r_i$, in the rescaled cosmological time the evolution of the hidden variable $r_i$ of node $i$ --- in both SPA and  $\widetilde{\mathrm{SPA}}$ --- is
\begin{equation}\label{eq:evol r}
r_i(t)=t_i+\frac{1-\beta}{\beta}t,
\end{equation}
where $t_i=\beta \ln i$ is the birth time of node $i$. The expected node degrees, however, evolve differently in SPA and $\widetilde{\mathrm{SPA}}$, as we show below.

\subsection{Evolution of the expected node degree in SPA}
\label{sec: kappa SPA}

The expected degree of node $j$ at network time $i>j$ is
\begin{equation}\label{eq:expdegSPA}
\begin{split}
\kappa_j(i)=&\sum_{s<j}p_{js}+\sum_{s>j}^i p_{sj}\\
\approx& \int_0^j \hspace{-1mm}\frac{ds}{1+e^{r_s(j)+r_j-R_j}}+
\int_j^i \hspace{-1mm}\frac{ds}{1+e^{r_j(s)+r_s-R_s}}.
\end{split}
\end{equation}
The first term is the expected degree of node $j$ upon its birth, and the integral is calculated in the same way as the integral in (\ref{eq:<k>birth}), and it equals to $m$. The second term is
\begin{equation}
\begin{split}
&\int_j^i \frac{ds}{1+e^{r_j(s)+r_s-R_s}}\approx
\int_j^i \frac{ds}{1+\frac{j^{\beta}s^{1-\beta}}{m(1-\beta)}}\\
&\approx \frac{m(1-\beta)}{j^{\beta}}\int_j^i\hspace{-1mm} s^{\beta-1}ds=\frac{m(1-\beta)}{\beta}\left(\left(\frac{i}{j}\right)^{\beta}-1\right),
\end{split}
\end{equation}
and, therefore,
\begin{equation}\label{eq:kj(i)_final}
\kappa_j(i)=\frac{m(1-\beta)}{\beta}\left(\frac{i}{j}\right)^{\beta} + \frac{m(2\beta-1)}{\beta}.
\end{equation}
This expression coincides exactly with the expression obtained in \cite{DoMeSa00} for the expected degree of a node in the sharp preferential attachment model, where the link attraction probability is a liner function of the node's exact degree (versus expected degree).

Finally, in the rescaled cosmological time,
\begin{equation}\label{eq:ki(t)_final}
\kappa_i(t)=\frac{m(1-\beta)}{\beta}e^{t-t_i}+\frac{m(2\beta-1)}{\beta}.
\end{equation}

\subsection{Evolution of the expected node degree in $\widetilde{\bf{SPA}}$}
\label{sec: kappa tSPA}

Similarly, the expected degree of node $j$ at network time $i>j$ is
\begin{equation}
\begin{split}
\kappa_j(i)&=\sum_{s<j}p_{js}^{\mathrm{ext}} + \sum_{s>j}^i p_{sj}^{\mathrm{ext}}+\sum_{s>j}^i\sum\limits_{\substack{a<s\\ a\neq j}}p_{aj}^{\mathrm{int}}(s).
\end{split}
\end{equation}
The first two terms are exactly the same as in SPA (\ref{eq:expdegSPA}) with $m$ replaced by $m_{\mathrm{ext}}$. The last term is the expected contribution to the $j$'s degree from internal links,
\begin{equation}
\begin{split}
&\sum_{s>j}^i\sum\limits_{\substack{a<s\\ a\neq j}}p_{aj}^{\mathrm{int}}(s)\approx \int_j^i\int_0^s \frac{dads}{1+e^{r_a(s)+r_j(s)-R^{\mathrm{int}}}}\\
&= \int_j^i\int_0^s \frac{dads}{1+\frac{a^{\beta}j^{\beta}s^{2(1-\beta)}}{m_{\mathrm{int}}(1-\beta)}}\\
&\approx\frac{m_{\mathrm{int}}(1-\beta)}{j^{\beta}}\int_j^i\int_0^sa^{-\beta}s^{-2(1-\beta)}dads\\
&=\frac{m_{\mathrm{int}}}{\beta}\left(\left(\frac{i}{j}\right)^{\beta}-1\right).
\end{split}
\end{equation}
Combining all terms together, we obtain
\begin{equation}\label{eq:kj(i)_tilda_final}\begin{split} \kappa_j(i)=m_{\mathrm{ext}}\left(\frac{i}{j}\right)^{\beta},
\end{split}
\end{equation}
and, in the rescaled cosmological time,
\begin{equation}\label{eq:ki(t)_tilda_final}
\kappa_i(t)=m_{\mathrm{ext}}e^{t-t_i}=\bar{k}(1-\beta)e^{t-t_i}.
\end{equation}
If $\beta=1/2$, then, as expected, the expressions for the expected node degrees in SPA and  $\widetilde{\mathrm{SPA}}$ become identical.

\subsection{Dynamic Hamiltonians}\label{sec:dynamic-hamiltonians}

We now have all the ingredients necessary to derive the Hamiltonian describing the dynamics of network growth in $\widetilde{\mathrm{SPA}}$. We want to find Hamiltonian $\widetilde{\mathcal{H}}$ such that its Hamilton's equations have (\ref{eq:evol r}) and (\ref{eq:ki(t)_tilda_final}) as the solutions.

Let $\widetilde{\mathcal{H}}_i(\kappa_i,r_i,t)$ be the energy contribution of node $i$ at time $t$ to the total  network Hamiltonian
\begin{equation}
\widetilde{\mathcal{H}}=\sum_{i=1}^N\widetilde{\mathcal{H}}_i(\kappa_i,r_i,t).
\end{equation}
Then Hamilton's equations for node $i$ are
\begin{equation}\label{eq: Ham i}
\dot{\kappa}_i=\frac{\partial\widetilde{\mathcal{H}}_i}{\partial r_i}
\hspace{5mm}\mbox{and}\hspace{5mm}
\dot{r}_i=-\frac{\partial\widetilde{\mathcal{H}}_i}{\partial \kappa_i}.
\end{equation}
Formally integrating these equations, and noting that
\begin{equation}
\dot{\kappa}_i=m_{\mathrm{ext}}e^{t-t_i} \hspace{5mm}\mbox{and}\hspace{5mm}\dot{r}_i=\frac{1-\beta}{\beta},
\end{equation}
we can write the solution in the following form
\begin{equation}
\widetilde{\mathcal{H}}_i(\kappa_i,r_i,t)=m_{\mathrm{ext}}e^{t-t_i}r_i-\frac{1-\beta}{\beta}\kappa_i+\xi_i(t),
\end{equation}
where $\xi_i(t)$ is some function of time $t$ and model parameters $\bar{k}$ and $\beta$.

Since $\xi_i(t)$ does not affect the equations of motion (\ref{eq: Ham i}), in principle, it can be chosen arbitrary.  Remarkably, $\xi_i(t)$ can be chosen in such way that it is the same for all nodes, and the resulting total Hamiltonian $\widetilde{\mathcal{H}}$ is identical to the ERGM Hamiltonian $H_{\widetilde{\mathrm{SPA}}}$ with node degrees replaced by their expected values. Indeed, let
\begin{equation}
\label{eq:xi}
\xi_i(t)=\frac{\bar{k}}{2}\left(\frac{2(1-\beta)}{\beta}-\sigma-\frac{t}{\beta}\right).
\end{equation}
Now, consider a network snapshot at time
\begin{equation}\label{eq:T}
T=\beta\ln N\gg1,
\end{equation}
with given (current) values of $\kappa_i$ and $r_i$. Using (\ref{eq:ki(t)_tilda_final}), (\ref{eq:xi}) and (\ref{eq:T}), we can rewrite the Hamiltonian of node $i$ as follows
\begin{equation}
\widetilde{\mathcal{H}}_i=\kappa_ir_i-\frac{1-\beta}{\beta}(\kappa_i-\bar{k})-\frac{\bar{k}(\ln N+\sigma)}{2}.
\end{equation}
The total Hamiltonian of the snapshot is then
\begin{equation}
\widetilde{\mathcal{H}}=\sum_{i=1}^N\kappa_ir_i-\frac{1-\beta}{\beta}\left(\sum_{i=1}^N\kappa_i-N\bar{k}\right)-\frac{N\bar{k}(\ln N +\sigma)}{2}.
\end{equation}
Since $\sum_{i=1}^N\kappa_i=N\bar{k}$, we  obtain that
\begin{equation}\label{eq:HSPAdyn}
\begin{split}
\widetilde{\mathcal{H}}=&\sum_{i=1}^N\kappa_ir_i-\frac{(\ln N +\sigma)}{2}\sum_{i=1}^N\kappa_i\\
=&\sum_{i=1}^N\kappa_ir_i-\frac{\ln\left(N\bar{k}(1-\beta)^2\right)}{2}\sum_{i=1}^N\kappa_i,
\end{split}
\end{equation}
which is exactly the expected ERGM Hamiltonian~(\ref{eq:HPAextint}). Note that since the ERGM Hamiltonian can be interpreted as the energy of a given network snapshot at some time $t$, the dynamic Hamiltonian with $\xi_i(t)$ in (\ref{eq:xi}) yields indeed the expected energy of the snapshot.

To obtain the dynamic SPA Hamiltonian, all we need to do is to set $\beta=1/2$ in the above derivations. The energy contribution of node $i$ is then
\begin{equation}
\mathcal{H}_i=\frac{\bar{k}}{2}e^{t-t_i}-\kappa_i+\bar{k}\left(1-\frac{1}{2}\ln\frac{\bar{k}}{4}-t\right),
\end{equation}
and the total Hamiltonian that describes the dynamics of growing networks in SPA is
\begin{equation}
\mathcal{H}=\sum_{i=1}^N\kappa_ir_i-\frac{\ln\left( N\bar{k}/4\right)}{2}\sum_{i=1}^N\kappa_i.
\end{equation}
As expected, this Hamiltonian is exactly the ERGM Hamiltonian of SPA (\ref{eq:HPA}) with the node degrees replaced by their expected values.

\section{Conclusion}
\label{sec:Conclusions}

We have studied the dynamics of networks growing according to preferential attachment, and obtained two important results. First, we have shown that soft preferential attachment can be casted as an equilibrium exponential random graph model,  nearly identical to the soft configuration model.
In other words, the ensemble of random graphs that preferential attachment generates is nearly identical to the equilibrium ensemble of random graphs with power-law degree distributions, meaning that preferential attachment and configuration model generate any
graph $G$ of any size $N$ with the same probability $\P(G)$.
In general, this result is important because equilibrium network models tend to be more amenable for analytic treatment. In particular, this result, for the first time to the best of our knowledge, provides an explicit expression $\mathbb{P}(G)\propto\exp[-H(G)]$ with Hamiltonian $H(G)$~(\ref{eq:HPA}) for the probability $\P(G)$ that preferential attachment generates any given network $G$. The knowledge of $\P(G)$ can be used, for example, for answering the question of how likely it is that a given real network has been grown according to preferential attachment. This question can now be answered by standard techniques, such as comparing the probabilities $\P(G)$ of the typical preferential attachment networks and the real network under study.  Another application is an alternative simpler method to generate preferential attachment networks, which has already been implemented and publicly released as a part of a more general software package that generates random hyperbolic graphs~\cite{Aldecoa2015Generator}.

Second, we have demonstrated that the growing dynamics of preferential attachment networks is Hamiltonian. Remarkably, the Hamiltonian $\mathcal{H}$~(\ref{eq:HSPAdyn}) that defines the equations of motion~(\ref{eq:evol r},\ref{eq:ki(t)_tilda_final}) describing network dynamics is nearly identical to the ERG Hamiltonian $H$~(\ref{eq:HPAextint}). The only difference between the two is that the exact node degrees in $H$ are replaced by their expected values in $\mathcal{H}$.

These results may appear quite surprising at the first glance, but there is
an intuitive explanation. On the one hand, the equilibrium Hamiltonian $H(G)$ in the soft configuration model is
the energy of graph $G$ in the Boltzmann distribution~$\P(G)$ of this exponential random graph model.
This energy is the sum of energies of all edges in graph $G$, and
one can check that the energy of edge $\{ij\}$ is simply the sum of $i$'s and $j$'s Lagrange multipliers $r_i+r_j$.
On the other hand, as shown in Section~\ref{sec:PAasERG}, soft preferential attachment, at
each time~$t$, is also a similar exponential random graph model, with hidden variables $r_i$ playing the
role of Lagrange multipliers, and the energy of edge $\{ij\}$ at
time $t$ is also $r_i(t) + r_j(t)$.

In simpler terms, the reason behind this equivalence is quite
physical: both the dynamic Hamiltonian in preferential attachment and the equilibrium Hamiltonian in the configuration model are
system energies, albeit the established equivalence between the
growing and equilibrium representations of the same system is
slightly atypical in physics~\cite{KrOs13}.

Very few real networks can be adequately modeled as random graphs in the configuration model, which suggests that some additional terms must be added to the Hamiltonian to adequately describe the dynamics of different real networks.In this context, it is an interesting observation that the ERG ensemble that we found to be equivalent to preferential attachment is a degenerate case of the more general geometric network ensembles, which can be considered as a Fermi gas in a hyperbolic space, whose symmetry group is the Lorentz group~\cite{KrPaKi10,PKSBK12}. This observation calls for extending the developed canonical formalism for network analysis to this more general geometric case with non-degenerate symmetries. This extension is a highly non-trivial task for a number of technical reasons, but if successful, it may shed some light on the second question we raised in the introduction, concerning small-scale dynamics of networks.

We have  shown that preferential attachment can be formulated within the canonical formalism, in which the time evolution of a system is described by Hamilton's equations $\dot{q}=\partial\mathcal{H}/\partial p$ and $\dot{p}=-\partial\mathcal{H}/\partial q$.
The traditional application of the Hamiltonian formalism  in mathematical physics~\cite{Arnold-Book} deals with the following direct problem: given a Hamiltonian $\mathcal{H}$, which in most cases is the energy of the system, find the solution of the corresponding dynamical equations of motion. However, in physics history, the problem has almost always been inverse: first, chronologically, the equations of motion are found by some other, usually experimental methods, and only much later it is recognized by theoreticians that these equations are solutions of some Hamiltonian or Lagrangian systems defined by their symmetry groups. This was the case in most physics theories, from classical mechanics~\cite{Arnold-Book} to general relativity~\cite{Wald2010}. Our understanding of network dynamics seems to have been driven along a similar historic path. First preferential attachment was suggested as a likely mechanism responsible for the emergence of scale-free degree distributions~\cite{BarAlb99,KraReLe00,DoMeSa00}, experimentally validated for many real networks~\cite{Ne01,BaJeNe02,VaPaVe02a,JeNeBa03}.
And only fifteen years later have we recognized that the preferential attachment dynamics~(\ref{eq:evol r},\ref{eq:ki(t)_tilda_final}) is Hamiltonian~(\ref{eq:HSPAdyn}).

We emphasize however that these results hold only for the soft versions of preferential attachment and configuration model. The difference between the soft configuration model with a fixed expected scale-free degree sequence and the configuration model in which the expected degree sequence is sampled for each graph from a fixed scale-free distribution has been recently quantified in~\cite{AnKr14}. This difference is well-behaved, in the sense that the entropy distribution in the latter ensemble is self-averaging, meaning that its relative variance vanishes in the thermodynamic limit. However, it is known that the soft (canonical) and sharp (microcanonical) configuration models are different even in the thermodynamic limit---the ensemble distributions do not converge in the limit~\cite{AnBi09,Squartini2015Equivalence}. To the best of our knowledge, there are no results of this sort concerning the difference between the soft and sharp versions of preferential attachment, but one could expect them to be different as well. Therefore the existence of any connections between sharp configuration model and sharp preferential attachment, and the possibility to formulate the latter within the canonical approach, remain to be open questions.

\begin{acknowledgments}
We thank Paul Krapivsky for useful discussions.
This work was supported by DARPA grant No.\ HR0011-12-1-0012; NSF grants No.\ CNS-1344289, CNS-1442999,
CNS-0964236, CNS-1441828, CNS-1039646, and CNS-1345286; by Cisco Systems; and by a Marie Curie International Reintegration Grant within the 7th European Community Framework Programme.
\end{acknowledgments}

\appendix
\renewcommand{\thesubsection}{A.\arabic{subsection}}
\renewcommand{\theequation}{A.\arabic{equation}}

\section*{Appendix}

\subsection{SPA, $\widetilde{\bf{SPA}}$, and SCM$^+$ as $\beta\rightarrow1$ ($\gamma\rightarrow2$)}\label{A1}

All three models --- SPA, $\widetilde{\mathrm{SPA}}$, and SCM$^+$ ---  have singularities at $\beta=1$. In this Appendix, we investigate to what models they degenerate in the limit $\beta\rightarrow1$.

The SPA model has a well-defined limit. Indeed, since
\begin{align}
R_i&\rightarrow r_i-\ln\frac{r_i}{m}\hspace{3mm} \mbox{and}\\
p_{ij}&\rightarrow\frac{1}{1+e^{r_j+\ln r_i-\ln m}}=\frac{1}{1+\frac{j\ln i}{m}},
\end{align}
as $\beta\rightarrow1$, SPA converges to the following simple model. To generate a network of size $N$ with average degree $\bar{k}$ and power-law exponent $\gamma=2$, for each new node $i=1,\ldots,N$, connect node $i$ to each existing node $j<i$ with probability
\begin{equation}
p_{ij}=\frac{1}{1+\frac{j\ln i}{m}}, \hspace{3mm} m=\frac{\bar{k}}{2}.
\end{equation}

\begin{figure}
	\centerline{\includegraphics[width=87mm]{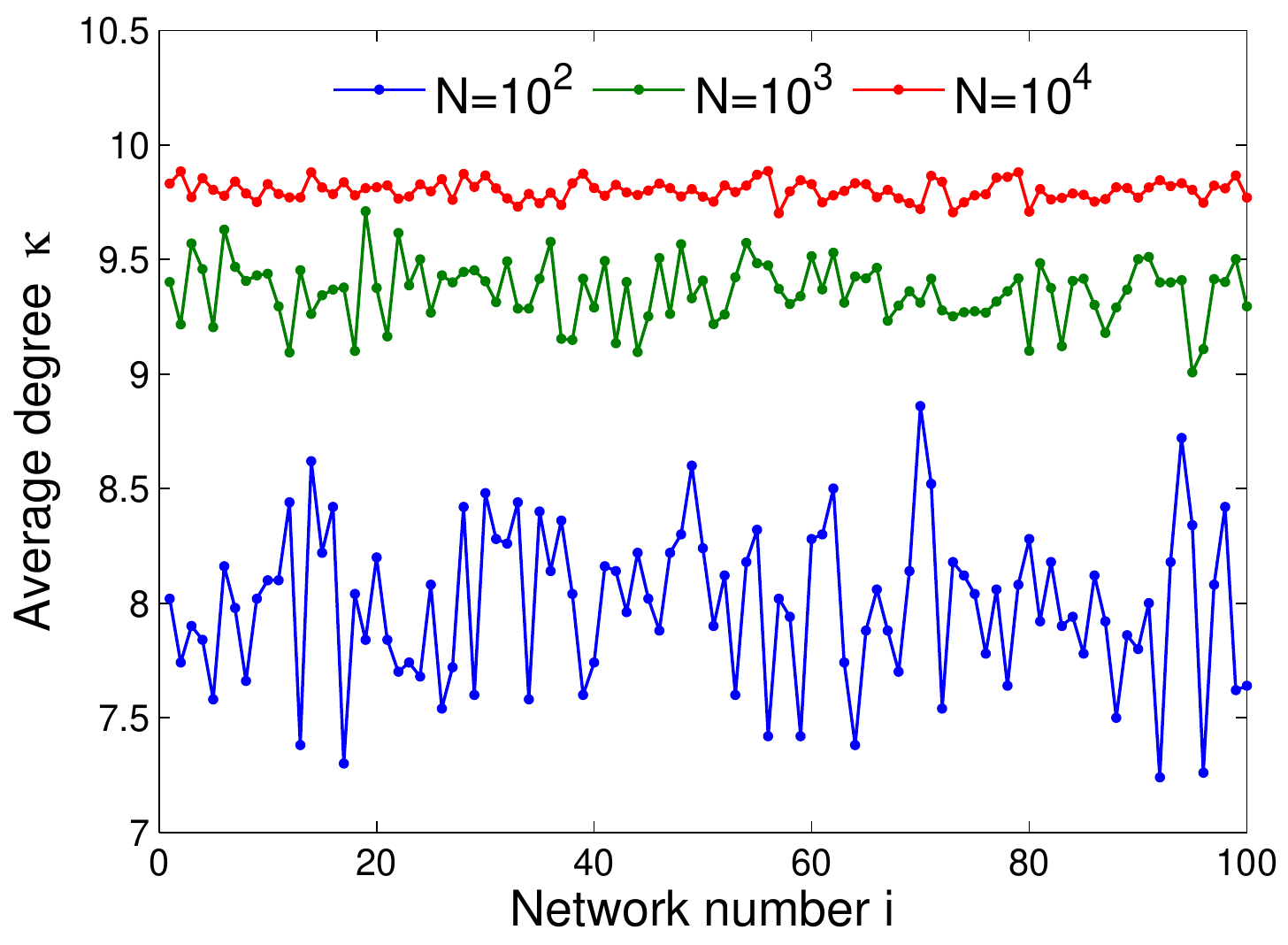}}
	\caption{\small \textbf{Average degree in SPA networks with $\beta=1$ ($\gamma=2$).}  For $N=10^2, 10^3,$ and $10^4$, the plot shows the fluctuation of the average degree in 100 independent networks generated by the limiting ($\beta=1$) SPA model with $\bar{k}=10$. As expected, the lager the network size $N$, the more accurate the approximation (\ref{eq:expevedegPAbeta=1}). } \label{Fig3}
\end{figure}

In what follows, we show that the average degree in large networks generated by this limiting model is indeed $\bar{k}$. At time $N$, the expected degree of node $i$ is
\begin{equation}\label{eq:expdegN}
\kappa_i(N)=\sum_{j<i}p_{ij}+\sum_{j>i}^Np_{ji}.
\end{equation}
The first sum is the expected contribution to the $i$'s degree from older nodes,
\begin{equation}
\sum_{j<i}p_{ij}\approx \frac{m}{\ln i}\sum_{j=1}^i\frac{1}{j}=\frac{m}{\ln i}H_i\approx m,
\end{equation}
where $H_i=\sum_{j=1}^i\frac{1}{j}\approx\ln i$ is the $i^{\mathrm{th}}$ harmonic number. The second sum in (\ref{eq:expdegN}) is the expected contribution to the degree of node $i$ from younger nodes,
\begin{equation}
\sum_{j>i}^Np_{ji}\approx \frac{m}{i}\int_i^N\frac{dj}{\ln j}=\frac{m}{i}({\rm li}(N)-{\rm li}(i)),
\end{equation}
where ${\rm li}(x)$ is the logarithmic integral function. Therefore,
\begin{equation}
\kappa_i(N)=m+\frac{m}{i}({\rm li}(N)-{\rm li}(i)).
\end{equation}
The expected average degree in the network at time $N$ is then
\begin{equation}
\begin{split}
\kappa(N)=&\frac{1}{N}\sum_{i=1}^N \kappa_i(N)\\
=&m+\frac{m{\rm li}(N)H_N}{N}-\frac{m}{N}\sum_{i<N}\frac{{\rm li}(i)}{i}.
\end{split}
\end{equation}
Since ${\rm li}(N)\approx \frac{N}{\ln N}$ and
$H_N\approx\ln N$, the second term is approximately $m$. The last term can be approximated as follows
\begin{equation}
\sum_{i<N}\frac{{\rm li}(i)}{i}\approx\int_0^N
\frac{{\rm li}(i)di}{i}={\rm li}(N)\ln N - N.
\end{equation}
Therefore, we finally have
\begin{equation}\label{eq:expevedegPAbeta=1}
\begin{split}
\kappa(N)=&2m-\frac{m}{N}({\rm li}(N)\ln N - N)\\
=&3m-\frac{m{\rm li}(N)\ln N}{N}\approx 2m=\bar{k}.\\
\end{split}
\end{equation}
Figure~\ref{Fig3} illustrates how this approximation becomes more accurate as the network size increases.

The $\widetilde{\mathrm{SPA}}$ model completely degenerates as $\beta\rightarrow1$. Namely,
$R_i^{\mathrm{ext}}\rightarrow-\infty$ and  
$R^{\mathrm{int}}\rightarrow-\infty$,
and, therefore,  $p_{ij}^{\mathrm{ext}}\rightarrow0$ and $p_{ab}^{\mathrm{int}}(i)\rightarrow0$. This means that the limiting model generates networks with no links. Remarkably, even in the limit $\beta\rightarrow1$, $\widetilde{\mathrm{SPA}}$ remains strongly equivalent to SCM$^+$. To prove this, we need to show that in this limit SCM$^+$ also generates networks without links.

The connection probability in SCM$^+$ is
\begin{equation}
p_{ij}=\frac{1}{1+e^{r_i^++r_j^+-R_{\mathrm{SCM}^+}}},
\end{equation}
where $r_i^+=r_i+\ln\left(\bar{k}(1-\beta)^2\right)$, $r_i\sim\rho_{\mathrm{SCM}}(r)$, and $R_{\mathrm{SCM}^+}=\ln\left(N\bar{k}(1-\beta)^2\right)$.
Since
\begin{equation}
e^{r_i^++r_j^+-R_{\mathrm{SCM}^+}}=\frac{e^{r_i}e^{r_j}}{N}\bar{k}(1-\beta)^2,
\end{equation}
and the expected value
\begin{equation}
\begin{split}
\langle e^{r_i} \rangle=&\int_0^{R_{\mathrm{SCM}}}e^r\rho_{\mathrm{SCM}}(r)dr\\
\approx&\int_0^{R_{\mathrm{SCM}}}e^re^{r-R_{\mathrm{SCM}}}dr\\
=&\frac{e^{R_{\mathrm{SCM}}}-e^{-R_{\mathrm{SCM}}}}{2}\rightarrow\frac{N}{2\bar{k}(1-\beta)^2},
\end{split}
\end{equation}
we have that
\begin{equation}
e^{r_i^++r_j^+-R_{\mathrm{SCM}^+}}\approx\frac{N}{4\bar{k}(1-\beta)^2}\rightarrow\infty.
\end{equation}
This means that the connection probability in SCM$^+$ converges to zero, $p_{ij}\rightarrow 0$, as $\beta\rightarrow1$, and therefore, even in this degenerate regime $\widetilde{\mathrm{SPA}}$ and SCM$^+$ are strongly equivalent.

\subsection{Accuracy of the classical limit approximation for the Fermi-Dirac  distribution in SCM$^+$}\label{A2}

\begin{figure}
	\centerline{\includegraphics[width=85mm]{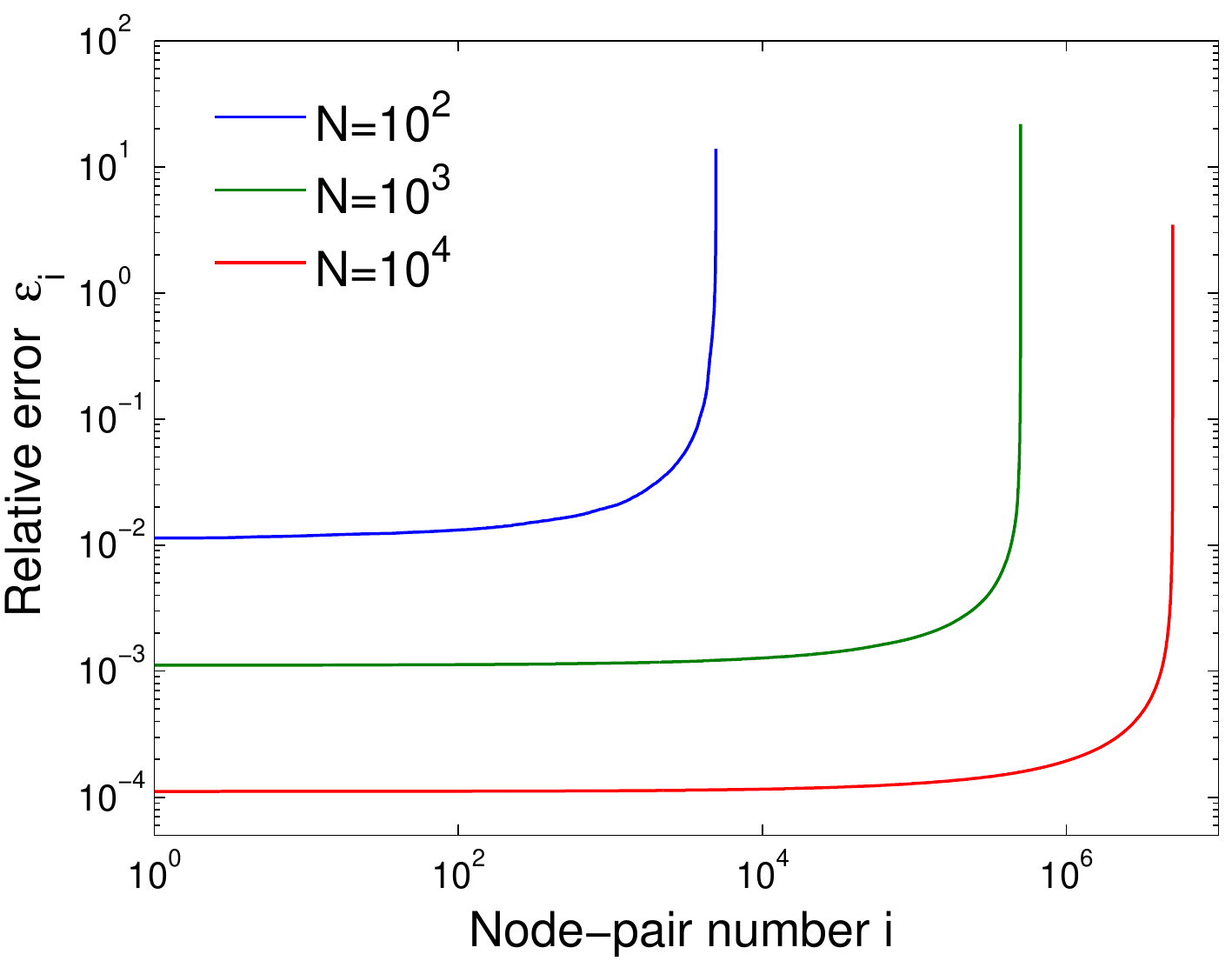}}
	\caption{\small \textbf{Accuracy of the classical limit approximation.} For $N=10^2, 10^3,$ and $10^4$, the percentage of node-pairs with the relative error larger than $5\%$ ($1\%$) is, respectively, 40.5\% (100\%), 1.8\% (13.6\%), and 0.08\% (1\%).  } \label{Fig4}
\end{figure}

In Section~\ref{sec:tSPA=SCM+}, we used the classical limit for the Fermi-Dirac distribution in the SCM$^+$ model
\begin{equation}\label{eq:approx}
p_{\mathrm{SCM}^+}(r,\acute{r})\approx\hat{p}_{\mathrm{SCM}^+}(r,\acute{r})=\frac{\bar{k}(1-\beta)^2N}{e^{r+\acute{r}}}.
\end{equation}
Here we show with simulations that this approximation is very accurate in large networks. As an example, we consider networks with $\bar{k}=10$ and $\gamma=2.5$. First, we generate $N$ hidden variables $r$ from distribution $\rho_{\mathrm{SCM}^+}(r)$, and then, for each of the ${N\choose 2}$ pairs of nodes with hidden variables $r$ and $\acute{r}$, we compute the relative error of the connection probability approximation (\ref{eq:approx})
\begin{equation}
\varepsilon=\frac{|p_{\mathrm{SCM}^+}(r,\acute{r})-\hat{p}_{\mathrm{SCM}^+}(r,\acute{r})|}{p_{\mathrm{SCM}^+}(r,\acute{r})}.
\end{equation}
Figure~\ref{Fig4} shows the relative errors sorted in the increasing  order for network sizes $N=10^2, 10^3,$ and $10^4$. As expected, the larger the network, the smaller the classical limit approximation error.

\bibliography{bib}

\end{document}